\documentclass{article}

\usepackage{graphicx}
\usepackage{multirow}
\usepackage{booktabs}
\usepackage{natbib}

\usepackage{amsmath,amssymb,amsfonts, appendix}
\usepackage{mathrsfs}
\usepackage{bm}
\usepackage{placeins}

\usepackage{xcolor}
\usepackage{tikz}
\usepackage{comment}
\usepackage{hyperref, url}
\usetikzlibrary{arrows.meta,positioning}

\newtheorem{theorem}{Theorem}
\newtheorem{proof}{Proof}
\newtheorem{proposition}[theorem]{Proposition}
\newtheorem{lemma}[theorem]{Lemma}
\newtheorem{example}{Example}

\newtheorem{definition}{Definition}
\providecommand{\keywords}[1]
{
  \small	
  \textbf{\textit{Keywords---}} #1
}

\title{The Spoiler Effect in Multiwinner Ranked-Choice
Elections: Implications for Proportional
Representation}
\author{
  Brian Brubach\\
  Department of Computer Science, Wellesley College \\
    \and
  Matthew I. Jones \\
  Department of Mathematics, Colby College\\
  \and 
 David McCune\\
 Department of Mathematics, William Jewell College\\
 \and
 Jennifer Wilson\\
 Department of Natural Sciences and Mathematics, The New School}

\begin{document}
\maketitle
\abstract{
We investigate the spoiler effect in multiwinner ranked-choice elections, in which removing a losing candidate changes the set of winners. Using ballot data from 999 Scottish local government elections, we compare the susceptibility of several multiwinner voting rules to spoiler effects and examine the relationship between spoiler susceptibility and proportional representation. We find that spoiler susceptibility varies considerably across voting methods but is not closely associated with whether a method is proportional or majoritarian. In particular, the Scottish version of single transferable vote (STV) exhibits spoiler effects in 49 elections, while Meek STV exhibits spoiler effects in only 24. Both methods are relatively resistant to spoilers compared with most of the other rules considered. We also investigate the effects of introducing cloned candidates, examine the characteristics of spoiler candidates, and establish theoretical worst-case results for several voting methods. Finally, we examine whether spoiler effects under STV compromise proportional representation, finding that the removal of a spoiler candidate does not change whether the winning committees satisfy three standard proportionality axioms. Our results suggest that spoiler effects under STV are relatively uncommon and do not necessarily indicate a failure of proportional representation.
}

\keywords{Single transferable vote, Spoiler effect,
Ranked-choice voting, Proportional representation,
Scottish local government elections}


\section{Introduction}

Multiwinner voting methods are used in a wide variety of settings, including elections of multi-seat districts or city councils, the selection of shortlists of job applicants, and the choice of webpages returned by a search engine. The social choice literature often takes an axiomatic approach to evaluating such rules, articulating fairness criteria, identifying which methods satisfy them, and estimating how frequently violations occur. In this article we focus on one fairness criterion--spoiler-proofness--and compare how frequently different multiwinner voting methods exhibit the spoiler effect. Our primary voting rule of interest is the single transferable vote (STV), the most commonly used method for ranked-choice multiwinner political elections. We examine other voting methods to provide context for STV's performance. Our general finding is that STV is quite resistant to spoilers in practice, although some other methods are more resistant.

Our study is motivated in part by the fact that the single-winner implementation of STV (known as instant runoff voting, the Hare method, or colloquially as ``ranked-choice voting'') is widely touted as being \emph{spoiler-proof} or virtually spoiler-proof. Previous work \citep{MW} has largely supported this claim; we show that the multiwinner picture is less clear. Under STV, the presence or absence of a losing candidate can alter not only the order of elimination, but also quota values and surplus transfers, creating additional opportunities for the spoiler effect.

We also examine whether the presence of a spoiler affects the proportionality of an STV election. Multiwinner voting methods are generally characterized as \emph{majoritarian} (electing candidates with broad overall support; see \citet{EFSS, FSST}) or \emph{proportional} (electing candidates representing different blocs of voters). STV falls firmly in the proportional camp, satisfying proportionality for solid coalitions \citep{D}. Thus, it is of particular interest to analyze how the composition of a winning committee changes with the presence or absence of a spoiler candidate.

The formal study of multiwinner voting rules is a comparatively recent development within social choice theory. While the modern mathematical study of single-winner voting dates to the eighteenth-century work of Borda and Condorcet, the axiomatic analysis of multiwinner methods did not become an active area of research until the late twentieth century, with foundational contributions by \citet{Debord}, \citet{D}, \citet{FM}, and \citet{Fishburn}. Since our work is motivated by STV, we focus on preference-based methods, in which voters submit possibly partial rankings of the candidates. Previous work in this setting has examined monotonicity properties \citep{AL, EFSS, FM, MGS, Staring}, generalizations of the single-winner Condorcet criterion \citep{AEFLS, ELS, Fishburn, Gehr, K, Rat}, and axioms of proportional representation \citep{AL, BP, D, Woodall}.

Previous work most closely related to ours falls into two categories. The first concerns spoiler effects in multiparty elections. \citet{BSSS} and \citet{K18} examine ``spoiler parties'' in an apportionment-based framework. The second concerns clone independence, first studied in the single-winner setting by \citet{Ti87}. \citet{Woodall} formulated several versions of clone independence for multiwinner elections and examined their relationship to STV. More recently, \citet{NR} studied the computational complexity of cloning candidates under several multiwinner rules, including STV, while \citet{D26} studies approximate clones, including an empirical analysis using the same Scottish election dataset we analyze.

The spoiler effect has also received considerable attention in the single-winner setting. Depending on its definition, related work dates to Arrow's classical independence of irrelevant alternatives (IIA) axiom \citep{Arrow}. For example, \citet{Miller} discusses the spoiler effect in relation to the consistency property and IIA. Our work primarily extends the analysis of \citet{MW}, who examine 170 municipal and statewide elections in the United States to compare the susceptibility of instant runoff voting and plurality to the spoiler effect. They also investigate how frequently spoilers are ``weak'' candidates, defined as Condorcet losers (see also the supplementary note by \citet{Miller_note}).

We extend this analysis to the multiwinner setting, comparing STV with a variety of preference-based voting methods, both proportional and majoritarian. Using data from 999 Scottish local government elections conducted under STV, we measure how frequently each method exhibits the spoiler effect and how often spoilers are weak according to two natural measures: plurality losers and top-$k$ losers (candidates appearing in the fewest voters' top-$k$ rankings). We also investigate whether susceptibility to spoilers is related to a method's classification as proportional or majoritarian.

Our main findings are:

\begin{itemize}
\item STV is quite resistant to the spoiler effect in practice. In 999 Scottish elections, Scottish STV exhibits a spoiler effect in 49; the number drops to 24 for Meek STV.
\item Resistance to the spoiler effect has little relationship to whether a voting method is proportional or majoritarian.
\item Spoiler effects under STV exhibit a mix of vote-splitting and non-vote-splitting behavior, with a slight tendency toward the former. Furthermore, STV is highly resistant to strategically problematic changes in the winning committee resulting from the addition of clone candidates.
\item There is a large gap between STV's empirical behavior and its theoretical worst-case behavior.
\item Under Scottish STV, we find no spoiler instance in which the resulting committee change creates a detectable loss of proportional representation under several standard proportionality axioms.
\end{itemize}

The paper is structured as follows. Section \ref{prelims} provides relevant definitions. Section \ref{section:scottish_data} describes the Scottish dataset, while Section \ref{section:results_from_scottish_data} presents baseline results. Section \ref{section:clones} investigates clone candidates and vote-splitting, Section \ref{section:PR} examines the implications for proportional representation, and Section \ref{section:worst_case} considers theoretical worst-case outcomes. Section \ref{section:conclusion} concludes.

\section{Preliminaries}\label{prelims}
We  consider only elections in which voters submit rankings over candidates. Hence we define an election as a triple $\mathcal{E}=(\mathcal C,\mathcal{P},k)$ where 
$\mathcal C$ is a set of $m$ candidates and $\mathcal P$ is a preference profile of $n$ voters --- a collection of $ n$ strict, possibly  partial, rankings of the candidates. Furthermore, $k \leq m$ is the size of the winning committee. Candidates omitted from a ballot are treated as tied below all ranked candidates. A multiwinner voting rule $f$ assigns, for each election $(\mathcal C,\mathcal P,k)$ a set of winning committees $f(\mathcal C,\mathcal P, k)=\{\mathcal{W}_1,\ldots,\mathcal{W}_\alpha\},$
where each $\mathcal{W}_i\subseteq \mathcal C$ has size $k$. The voting rules we consider are generally \textit{resolute}---that is, they almost always produce a unique winning committee, with multiple winning committees arising only from infrequent ties. When the voting rule is clear, we denote the set of winning committees by $\{\mathcal{W}_1,\ldots,\mathcal{W}_\alpha\}$, or simply by $\mathcal{W}$ when the winning committee is unique.

 We define a \emph{spoiler candidate} to be a losing candidate $S$ that, by its removal, changes the winner set of the election. Throughout the article we use $S$ to denote such a candidate. More precisely, let $\mathcal{C}^{-S}$ be the candidate set with $S$ removed and let $\mathcal{P}^{-S}$ denote the ballot profile obtained by removing $S$ from every ballot in $\mathcal P$ and shifting candidates ``up'' the rankings accordingly after $S$ is removed (in particular, ballots which rank only $S$ are completely removed in $\mathcal{P}^{-S}$). Then $S$ is a \emph{spoiler candidate} if $S$ does not belong to any winning committee in $f(\mathcal C,\mathcal P, k)$ and $f(\mathcal C,\mathcal P, k)\neq f(\mathcal{C}^{-S},\mathcal{P}^{-S},k).$ For brevity, if $ f(\mathcal{C}^{-S},\mathcal{P}^{-S},k)$ contains only one committee then we use $\mathcal{W}^{-S}$ to denote the winning committee when the spoiler $S$ is removed. An election exhibits the \emph{spoiler effect} under a voting rule if it has at least one spoiler candidate. If $\mathcal{W}$ and $\mathcal{W}^{-S}$ differ by only a single candidate, we let $R$ denote the candidate in $\mathcal{W}$ but not in $\mathcal{W}^{-S}$ ($R$ is the ``real winner'') and $A$ denote the candidate in $\mathcal{W}^{-S}$ but not $\mathcal{W}$ (the ``alternate winner'').


A voting method that is more likely to exhibit the spoiler effect  or to be sensitive to clones is less desirable than one that isn't, particularly if the spoilers  tend to be weak in some sense. (We prefer to avoid weak spoiler candidates because they suggest that the election outcome may be affected by the presence or absence of a candidate who in some sense is irrelevant to the election.)  In this way, being susceptible to spoilers is similar to failing Arrow's famous IIA criterion. As with the IIA criterion, spoiler-proofness is virtually impossible for reasonable voting methods to achieve (see  \ref{sec:spoilers_inevitable}). However, some voting methods are much more susceptible to the spoiler effect than others.

Since spoiler candidates are considered especially egregious when they have little chance of being among the winners, we identify  two notions of weakness. The first is the {\it plurality loser}, which is the candidate with the fewest first-place votes. The second is the {\it top$-k$ loser}, which is the candidate who receives the fewest top-$k$ rankings across all ballots. As part of our results we report if a spoiler candidate is the plurality loser or the top-$k$ loser. Both notions identify candidates who are comparatively weak according to natural measures of electoral support, making spoiler effects involving these candidates particularly noteworthy.

\subsection{Multiwinner voting rules}

In this section, we briefly define the voting methods which we analyze in the paper. A fuller description is available in   \ref{voting_rules_app}. Because ties are rare in our work, we do not include a discussion of how to break ties as part of our definitions.

\emph{Single transferable vote}  (STV)  refers to a family of related voting rules. In this article we consider two versions of STV: the implementation used in Scottish local government elections (\emph{Scottish STV}) and the more mathematically sophisticated \emph{Meek STV} \cite{M94}, which is used for some local elections in New Zealand. Both versions use the {\it integer Droop quota} $q=\lfloor \frac{n}{k+1} \rfloor+1$ and  transfer of fractional votes.  The election proceeds in a series of rounds, in each of which either a candidate is eliminated or one or more candidates is selected to be part of the winning committee.   In each round, the number of first-place votes for each candidate is determined. Any candidate receiving at least $q$ votes is selected as a winner and the number of votes over the quota they receive (referred to as their {\it surplus}), is transferred to the eligible candidates next on the voters' ballots in proportion to the  votes those candidates receive.  
If no candidate receives at least $q$ votes then the candidate with the fewest first-place votes is eliminated. This process continues until either $k$ candidates have been selected or until some number $k^\prime<k$ candidates  have been selected and  only $k-k^\prime$ candidates remain who are then selected. A complete description of the STV ballot counting process as implemented in Scottish local government elections can be found at \url{https://www.legislation.gov.uk/sdsi/2007/0110714245}.

Meek STV is similar to Scottish STV except that under Meek, the quota is adjusted to reflect the number of currently active ballots, and candidates already elected can continue to receive vote transfers.  See \cite{HWW} for complete details of the Meek algorithm.

We also study the voting methods of bloc voting (Bloc), $k$-Borda, sequential ranked-choice voting (SRCV), single non-transferable vote (SNTV), Chamberlin-Courant (Cham-Cour), and the expanding approvals rule (EAR). Definitions of these methods are in  \ref{voting_rules_app}.

The three methods, $k$-Borda, Cham-Cour and EAR require interpretation when voters submit partial preferences.  (The remaining methods have natural adaptations for this setting.)  Following \cite{BFLR}, we use two models of handling partial ballots: an \emph{optimistic} and a \emph{pessimistic} model.  Under the optimistic model (OM), all candidates not appearing on a ballot receive a number of points equal to the next open rank on the ballot. Under the pessimistic model (PM), all candidates not appearing on a ballot receive zero points from that ballot. These models extend analogously to assigning points for sets of candidates under Cham-Cour, and to defining levels of approval at each round under EAR (see  \cite{GJM} for a fuller discussion).
Thus, we have OM and PM versions for the  methods of Borda, Cham-Cour, and EAR.

Finally, we define two methods which have not been previously studied to our knowledge, but were identified because they are highly resistant to the spoiler effect in our real-world elections, performing the best among the large pool of methods considered.  

The \emph{top-k IRV} rule iteratively eliminates the candidate with the fewest first-place votes until only $k$ candidates remain, who are declared the winners. Equivalently, this method works like STV where quota has been set at $n$. 

The final rule is based on the notion of a Condorcet committee \cite{Gehr, Rat}. A subset of candidates $\mathcal{X} \subset \mathcal{C}$ is a \emph{Condorcet committee} if, for every $A\in \mathcal{X} $ and $B\notin \mathcal{X} $, more voters prefer $A$ to $B$ than prefer $B$ to $A$. The \emph{Minimax Condorcet Committee} (MM Cond Committee) method   selects the (unique) Condorcet committee of size $k$. If no such committee exists, it finds the smallest Condorcet committee $\mathcal{X}$ of size $k'\geq k$ and selects the $k$ candidates in $\mathcal X$ with the highest margin score $\operatorname{score}(A)=\min_{B\in \mathcal{C} \setminus \{A\}}\operatorname{PW}(A,B),
$ where $\operatorname{PW}(A,B)$ is the number of voters who prefer $A$ to $B$ minus the number who prefer $B$ to $A$. 

\subsection{Classifying voting rules} As part of our study we analyze if proportional methods tend to be more susceptible to the spoiler effect than non-proportional methods. To carry out this analysis, we must classify each rule as majoritarian, semi-proportional, or proportional. There is no single universally accepted definition separating these three categories, and thus we use these terms as broad design classifications. Majoritarian rules emphasize candidates with broad aggregate support, while proportional rules seek to represent distinct groups of voters in approximate proportion to their size. Following the electoral-systems literature, we use semi-proportional for rules that facilitate minority representation and can produce approximately proportional outcomes, but do not provide the proportionality guarantees associated with fully proportional rules. SNTV is a canonical example of such a rule \cite{LA, NK}.

Under these broad notions, we classify Bloc, Borda OM, Borda PM, SRCV, and MM Cond Committee as majoritarian. We classify all versions of EAR, STV, and Cham-Cour as proportional. EAR and STV are considered proportional  primarily because they satisfy Dummett's classical proportionality axiom \emph{proportionality for solid coalitions} (PSC) (see \cite{D}). Cham-Cour was designed to promote proportional representation and is commonly treated as a representation-focused or proportional rule, although it does not satisfy PSC. We therefore group it with the proportional methods, while noting that under a classification based strictly on formal proportionality guarantees it could instead be viewed as semi-proportional. We classify SNTV as semi-proportional.

Since top-$k$ IRV has not been previously studied,  its classification has not been established in the literature. We classify the method as semi-proportional, although its behavior in our dataset suggests that it could also be described as ``empirically proportional.'' This assessment is supported by several observations. First, none of the winning committees selected by top-$k$ IRV in our dataset (described below) violates PSC. Second, top-$k$ IRV selects the same winning committee as Scottish STV in 862 of the 1,070 multiwinner elections (80.6\%) in the full Scottish dataset.\footnote{This number, and those that follow, may vary slightly depending on how ties are broken during candidate elimination under top-$k$ IRV.} Of the remaining 208 elections, top-$k$ IRV produces a winning committee containing candidates from more parties than the Scottish STV committee in 96 cases, whereas STV provides broader party representation in only 18 cases. Thus, although top-$k$ IRV does not satisfy proportionality guarantees in general, its empirical behavior in our dataset closely resembles that of a proportional voting rule.

\section{The Scottish Dataset}\label{section:scottish_data}

We study ballot data from Scottish local-government STV elections held between 2007 and 2022. Scottish council areas are divided into multi-member wards, which have elected councilors using STV on a five-year cycle since 2007.  We have access to the ballot data for 1128 Scottish elections, most of which come from the 2012, 2017, and 2022 election cycles (most of the data from 2007 is unavailable), as well as a handful of off-cycle by-elections. The full dataset for non-by-elections  is available at \url{https://github.com/mggg/scot-elex}; the ballot data for the by-elections is available by request.

In the majority of wards, $k=3$ or $k=4$: 58 elections satisfy $k=1$, five satisfy $k=2$, 554 satisfy $k=3$, 508 satisfy $k=4$, and three satisfy $k=5$. Because the spoiler effect cannot occur under any voting method if $m=k+1$, not all 1128 elections are useful for our analysis at all times. Furthermore, our focus is the multiwinner setting, and thus the 58 single-winner elections are not included in our  analysis. There are 999 elections that satisfy $m>k+1$ and $k>1$; these are the elections we use for the majority of this study. Across these 999 elections there are a total of 3473 seats available and 7479 candidates running, meaning there are 4006 losing candidates in total who are potential spoiler candidates.

An important methodological limitation of our empirical results is that our ballot data come from elections conducted using Scottish STV, so we cannot know how voting behavior might differ under other rules. We expect ballot data to be broadly similar under other proportional methods such as Meek STV, because incentives for voters, candidates, and parties would remain largely comparable. In contrast, majoritarian systems create different strategic incentives for voters, candidates, and parties. Such incentives could substantially alter voting behavior and, consequently, the observed ballot data. As a result, conclusions drawn from the Scottish dataset are most applicable to proportional voting methods. Other studies which use the Scottish data to provide a comparative analysis of different voting methods include \cite{BBMP24}, \cite{MMLS24}, and \cite{GJM}, which examine methods such as EAR and SRCV.

\section{Spoiler Results from the Scottish Dataset}\label{section:results_from_scottish_data}

\begin{table}
    \begin{tabular}{r|p{12 mm}|p{12 mm}|p{12 mm}|p{12 mm}|p{12 mm}|p{12 mm}|p{12 mm}|p{12 mm}}
        Method & Elections with spoilers & Multiple spoilers & Multiple winner sets & Number of spoiler candidates & Multiple winner swap spoilers & Number of spoiled seats & Plurality loser spoilers & Top k loser spoilers \\
        \hline \hline 
        Bloc & 66 & 35 & 4 & 116 & 0 & 67 & 36 & 36\\
        $k$-Borda OM & 165 & 40 & 11 & 224 & 0 & 176 & 65 & 56\\
        $k$-Borda PM & 44 & 18 & 4 & 68 & 0 & 44 & 18 & 22\\
        SRCV & 27 & 0 & 0 & 27 & 0 & 27 & 1 & 4\\
        MM Cond Comm & 4 & 0 & 0 & 4 & 0 & 4 & 1 & 1\\
        SNTV & 105 & 46 & 29 & 159 & 0 & 108 & 27 & 25\\
        Top $k$ IRV & 6 & 0 & 0 & 6 & 0 & 6 & 0 & 0\\
        Cham-Cour OM & 102 & 43 & 5 & 171 & 0 & 105 & 49 & 38\\
        Cham-Cour PM & 64 & 33 & 1 & 120 & 3 & 67 & 47 & 40\\
        Scottish STV & 49 & 9 & 2 & 66 & 0 & 49 & 9 & 10\\
        Meek STV & 24 & 0 & 0 & 24 & 0 & 24 & 1 & 2\\
        EAR OM & 215 & 97 & 44 & 386 & 3 & 234 & 81 & 90\\
        EAR PM & 125 & 56 & 10 & 239 & 1 & 128 & 52 & 50
        
    \end{tabular}
    \caption{Spoiler results from the Scottish dataset (999 elections). Column 1 lists all the methods, loosely ordered by majoritarian methods and then proportional methods. Column 2 counts the number of elections that have at least one spoiler candidate. Column 3 counts the number of elections with more than one spoiler. Column 4 counts the elections where different spoiler candidates can lead to different winner sets. Column 5 counts the total number of spoiler candidates. Column 6 counts the number of spoilers whose removal changes more than one winner. Column 7 counts the number of candidates who won seats that could have lost their seat if a spoiler was removed (out of 3473). Columns 8 and 9 count the number of elections where the plurality loser and the top-$k$ loser are spoiler candidates, respectively.}
    \label{tab:results_table}
\end{table}

We begin by examining how frequently each voting method exhibits the spoiler effect and how strong the resulting spoiler candidates tend to be. Table \ref{tab:results_table} summarizes the baseline results across the 999 Scottish elections. A handful of elections produce ties under some methods; we exclude these cases from the table for these methods to avoid treating a tie resolved by candidate removal as a spoiler effect.

Interestingly, there is no apparent relationship between a method's proportionality categorization and its resistance to the spoiler effect. Of the primary methods, the two most resistant are Meek STV (proportional) and SRCV (majoritarian), and the two most susceptible methods are Expanding Approvals Rule OM (proportional) and Borda OM (majoritarian). Thus, proportionality and resistance to spoilers appear largely unrelated.

The strongest pattern in Table \ref{tab:results_table} can be seen by comparing the optimistic and pessimistic versions of Borda, Chamberlin-Courant, and EAR. For all three methods, the optimistic version is much more susceptible to spoilers than the pessimistic version. One plausible explanation is that the optimistic model gives greater weight to weak candidates who do not appear on many ballots, creating a more competitive landscape in which such candidates can affect the outcome.

It is worth emphasizing that the values here are raw counts and should be considered in the context of our dataset. Under Scottish STV, only 49 of the 3,473 elected candidates (1.4\%) would lose their seats upon the removal of some spoiler candidate.

The final two columns of Table \ref{tab:results_table} also indicate that a non-trivial fraction of the spoilers are weak---being either plurality or top-$k$ losers. The number of weak spoilers tends to be strongly correlated with the number of spoilers generally, except Borda PM which has twice as many weak spoilers as Scottish STV, despite having fewer elections with spoilers.

To investigate how weak a spoiler tends to be under a given voting method beyond simply plurality loser or top-$k$ loser, we measure each spoiler's relative strength compared to the other losing candidates for plurality and top-$k$ scores. Let $L$ denote the number of losing candidates in an election, and let
$r_i$ denote the rank of losing candidate $i$ according to the relevant
measure of strength, where $r_i=1$ denotes the strongest losing candidate and $r_i=L$ denotes the weakest. We define the \emph{relative
strength} of candidate $i$ as
\[
R_i=\frac{L-r_i}{L-1}.
\]
Thus, $R_i=1$ for the strongest losing candidate and $R_i=0$ for the weakest losing candidate, with intermediate ranks evenly spaced between
0 and 1. For example, if there are five losing candidates, their relative-strength values, from strongest to weakest, are $1,  \frac{3}{4},  \frac{1}{2}, \frac{1}{4},  0.$  Thus, regardless of the number of candidates in an election, a score of 0 denotes that the candidate is the weakest among the losing candidates and a score of 1 denotes that the candidate is the strongest.

\begin{figure}

    \includegraphics[width=150mm]{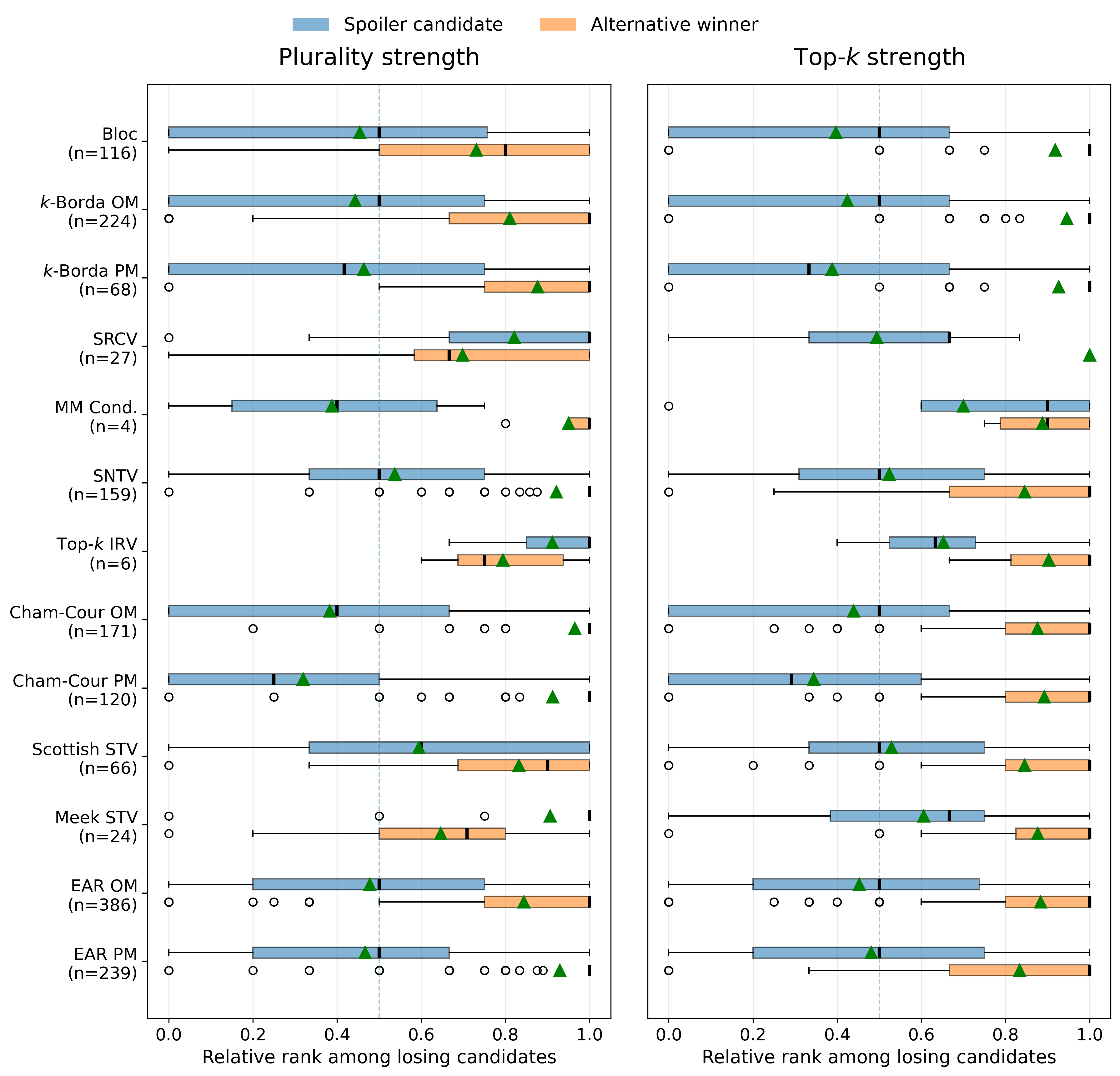}
    \caption{Relative strength of spoiler candidates and alternative winners under each voting method. Candidate strength is measured by rank among the losing candidates in the original election using plurality score (left) and top-$k$ score (right), with 0 denoting the weakest losing candidate and 1 the strongest. Triangles indicate means.}
    \label{figure:spoiler_relative_strength}
\end{figure}

The results are shown in Figure \ref{figure:spoiler_relative_strength}. When candidate strength is measured using plurality scores, Meek STV, SRCV, and top-$k$ IRV tend to exhibit the strongest spoilers, with the caveat that top-$k$ IRV returns only six spoiler candidates. In contrast, methods such as bloc, Borda, and Cham-Cour tend to exhibit weaker spoilers, although spoiler strength varies considerably under most methods. The overall pattern largely persists when strength is measured using top-$k$ scores. For example, under methods like Meek STV and SRCV, spoilers tend to be much stronger than for methods like bloc, Borda, or Cham Cour. Thus, methods that rarely exhibit spoilers do not necessarily do so because they eliminate only fringe candidates from consideration: under Meek STV and SRCV, the relatively rare spoilers that do occur tend to be comparatively strong losing candidates.

When spoiler effects occur, our analysis suggests a considerable degree of stability in the resulting winner sets under STV.  Even when an election has multiple spoiler candidates, removing them tends to result in the same winning committees. Comparing columns 3 and 4 of Table \ref{tab:results_table}, we see that most methods have drastically fewer elections with multiple winner sets than multiple spoilers. For example, there are 35 bloc elections with multiple spoilers, and for 31 of them, all spoilers result in the same winner set. Even for Borda (OM), under which one voter profile was susceptible to 7 different  spoilers (the 2017 election in the 4th Ward of the East Renfrewshire Council Area), the elimination of these spoilers resulted in only 3 alternate winning sets. The exceptions to this rule are SNTV and EAR, where many elections have multiple winner sets when different spoilers are removed.

When we consider the number of spoilers who change more than one winner on the committee, the results are even more suggestive of stability. Only Cham-Cour PM and EAR (OM and PM) have any spoilers that change multiple candidates. Of these spoilers, two of them (both in the Cham-Cour PM election for North Ayrshire 2012 Ward 5) replace 3 of the 4 winners on the committee, while the other 5 spoilers replace two candidates. This is in stark contrast to the worst case outcomes we discuss in Section \ref{section:worst_case}, where we see that spoilers can theoretically be pervasive and impactful; in practice, spoiler candidates have a limited effect on an election.


To conclude this section, we provide an example of an STV election where every losing candidate is a spoiler. This example helps illustrate how the removal of a candidate can change order of candidate eliminations and elections, resulting in a spoiler effect.

  \begin{example}\label{edinburgh}
  
  The 2022 election in Ward 5 of the City of Edinburgh Council unfolded as shown in the top of Table \ref{e-duns-table}, resulting in winners  Bandel, Mitchell, Nicolson and Osler.  Bandel wins a seat in the last round because  Nicolson surpasses quota by a wide margin when they are elected in the seventh round, allowing for a large transfer of surplus votes to Bandel.  If Herring is removed from the ballots then Nicolson does not have  nearly as many surplus votes to transfer when they are elected in the 4th (instead of the seventh) round, and Wood replaces Bandel in the winner set as shown in the middle of  Table \ref{e-duns-table}. The outcome is the same (with different numbers of surplus votes) if any other candidate besides Wood is removed.  If Wood is removed then Osler has a much larger initial surplus, more of which goes to Munro-Brian than to Bandel. As a result, after the first round of tabulation Munro-Brian has a lead over Bandel which Bandel cannot overcome, in contrast to the original election where the roles are reversed. Ultimately, Munro-Brian replaces Bandel in the winner set as shown in the bottom of Table \ref{e-duns-table}.

 \begin{table}
\footnotesize
\setlength{\tabcolsep}{4pt}
\begin{tabular}{c|c|c|c|c|c|c|c|c|c}
\multicolumn{10}{c}{\textbf{Actual Election}, Quota $=2684$}\\
\hline
Candidate & \multicolumn{9}{c}{Votes by Round}\\
\hline
Bandel (Grn)& 1714 & 1740.1&1740.1&1741.4&1751.5&1767.8&2221.6&2379.8&\textbf{2383.7}\\
Herring (Con)&853&863.1&863.1&867.1&889.3&&&&\\
Holden (SF)&96&97.4&98.4&109.4&&&&&\\
Laird (SL)&53&53.6&53.6&&&&&&\\
McNamara (Ind)&17&17.3&&&&&&&\\
Mitchell (Con)&1836&1877.5&1878.7&1883.8&1896.8&2643.6&\textbf{2767.6}&&\\
Munro-Brian (Lab)&1684&1713.0&1715.0&1721.0&1736.2&1755.3&&&\\
Nicolson (SNP)&2641&2657.3&2659.3&2663.3&2668.4&2683.4&\textbf{2936.8}&&\\
Osler (LD)&\textbf{3117}&&&&&&&&\\
Wood (LD)&1405&1700.6&1702.7&1711.7&1725.3&1765.1&2275.5&2303.0&2337.8\\
\hline
\end{tabular}

\vspace{.1 in}
\begin{tabular}{c|c|c|c|c|c|c|c|c}
\multicolumn{9}{c}{\textbf{Herring Removed}, Quota $=2677$}\\
\hline
Candidate & \multicolumn{8}{c}{Votes by Round}\\
\hline
Bandel (Grn)& 1728&1757.5&1757.5&1759.8&1760.4&1773.6&2331.8&2336.0\\
Holden (Con)&108&110.0&111.0&124.0&124.0&&&\\
Laird (SL)&60&60.8&60.8&&&&&\\
McNamara (Ind)&18&18.3&&&&&&\\
Mitchell (Con)&2530&2585.3&2586.5&2595.6&2595.7&2635.8&\textbf{2768.4}&\\
Munro-Brian (Lab)& 1698&1730.7&1732.7&1741.7&1741.9&1760.0&&\\
Nicolson (SNP)&2654&2672.1&2674.1&\textbf{2678.1}&&&&\\
Osler (LD)&\textbf{3168}&&&&&&&\\
Wood (LD)&1418&1754.0&1757.2&1767.2&1767.2&1783.9&2344.7&\textbf{2382.5}\\
\hline

\end{tabular}

\vspace{.1 in}

\begin{tabular}{c|c|c|c|c|c|c|c|c}
\multicolumn{9}{c}{\textbf{Wood Removed}, Quota $=2681$}\\
\hline
Candidate & \multicolumn{8}{c}{Votes by Round}\\
\hline
Bandel (Grn)&1726&1991.3&2036.3&2036.3&2041.6&2059.6&2082.9&2108.4 \\
Herring (Con)&886&1028.8&1029.3&1029.7&1035.5&1061.0&&\\
Holden (SF)&98&113.9&114.6&115.6&129.2&&&\\
Laird (SL)&53&64.1&64.4&65.8&&&&\\
McNamara (Ind)&17&21.4&21.5&&&&&\\
Mitchell (Con)&1845&2227.4&2228.8&2230.6&2238.2&2257.4&\textbf{3151.1}&\\
Munro-Brian (Lab)&1706&2054.3&2065.7&2068.7&2077.7&2101.9&2128.2&\textbf{2218.0}\\
Nicolson (SNP)&2645&\textbf{2756.1}&&&&&&\\
Osler (LD)&\textbf{4447}&&&&&&&\\
\hline

\end{tabular}

\caption{(Top) The votes in each round for candidates in the 2022 election in Ward 5 of the City of Edinburgh Council Area. (Middle) How the election would unfold if Herring were removed. (Bottom) How the election would unfold if Wood were removed.}
\label{e-duns-table}

\end{table}

 \end{example}
 
 Example \ref{edinburgh} demonstrates that multiwinner elections can exhibit different spoiler dynamics under STV than single-winner elections. If $k=1$ then it is not possible for all losing candidates to be spoiler candidates under STV, since the plurality loser cannot be a spoiler in this case.\footnote{If ties occur, then depending on how ties are broken and one's definition of spoiler effect in the case of ties, it is possible for the plurality loser to be a spoiler candidate.} Example \ref{edinburgh} therefore illustrates why the spoiler effect under multiwinner STV is qualitatively different from its single-winner counterpart: changes in quota, surplus transfers, and the timing of elections and eliminations create additional pathways through which losing candidates can affect the winning committee.

\section{Clones and the issue of ``vote-splitting''}\label{section:clones}

In this section we investigate the relationship between the spoiler effect and the closely-related notions of vote-splitting and candidate cloning. We approach this question from two directions. First, in Sections~\ref{subsec:votesplit} and \ref{subsec:stvcloserlook}, we use the spoiler effects identified in the Scottish data to ask whether spoiler candidates tend to resemble the candidates who would have won in their absence, as in a traditional vote-splitting scenario. Second, in Section~\ref{subsec:addclones}, we modify the ballot profiles directly by introducing artificial clones of existing candidates and examine how frequently these additions alter the winning committee. Together, these analyses assess the extent to which spoiler effects are driven by the presence of similar candidates.

\subsection{Vote-splitting}
\label{subsec:votesplit}

In the popular discourse concerning ranked-choice voting and the spoiler effect, the terms ``spoiler effect'' and ``vote-splitting'' are often used interchangeably. One reason for this is their frequent co-occurrence under plurality voting (the method of SNTV with $k=1$). Discussions of the spoiler effect under plurality often involve examples such as that shown in Table~\ref{votesplitting} in which Candidate $R$ is the plurality winner, but only because $A$ and $S$ ``split the vote''. If $S$ were not in the election, $A$ would presumably win easily. In this type of plurality spoiler event, $S$ can be a fringe candidate with very few first place votes or a popular candidate with more first place votes than $A$. Such examples motivate Tideman's definition of ``clone candidates'' \cite{Ti87} and its relationship to the spoiler effect. In this example,  $A$ and $S$ are clones because they appear in adjacent rankings on every ballot. See Section \ref{subsec:addclones} for further discussion of clones.

\begin{table}[tbh]
\centering

\begin{tabular}{l|ccc}
Num Voters & 100 & 90 & 40\\
\hline
1st choice & $R$ & $A$ & $S$\\
2nd choice &$A$ & $S$ & $A$\\
3rd choice & $S$ & $R$ & $R$\\
\end{tabular} 

\caption{Example of a spoiler under plurality due to vote-splitting}
\label{votesplitting}
\end{table}

In this section, we investigate the extent to which the spoiler effect can be attributed to vote-splitting under different multiwinner voting rules. We use the 999 Scottish elections to examine if spoiler candidates are similar to the candidate they prevent from winning. For each voting method, we identify all candidate triples of the form $(R,A,S)$ where $R$ is a member of the winning committee and $A$ is an alternative winner who would win a seat if the spoiler $S$ were removed from the election. Since only a handful of spoiler candidates across all methods caused the winner set to change by more than one candidate, we ignore these rare outlier cases. 


To assess whether the spoiler $S$ is splitting the vote with the alternative winner $A$, we use two related approaches. 
The first is to ask whether the voters ranking $S$ first prefer $A$ or $R$. We do this by checking who would win a head-to-head race between $A$ and $R$ using only the ballots which ranked $S$ first. When $A$ is preferred, this resembles the type of spoiler event captured in Table~\ref{votesplitting}. On the other hand, when $R$ is preferred, this reveals an interesting event in which the voters ranking the spoiler $S$ first actually prefer the ``spoiled'' outcome in which $R$ wins, and the spoiler effect is not due to the typical vote-splitting scenario. 
The second approach is to use the notion of strong similarity under the Mutual Preference Distance measure of candidate similarity introduced in \cite{Brubach}. 
Here, we say two candidates $C$ and $C'$ are \emph{strongly similar} if and only if the voters ranking $C$ first prefer $C'$ over all other candidates (besides $C$) and the voters ranking $C'$ first prefer $C$ over all other candidates (besides $C'$). 
As such, strong similarity is a symmetric relation, and a given candidate can be strongly similar to at most one other candidate. 
For each method, we check what percentage of spoiler events have strongly similar candidates $S$ and $A$ and what percentage have strongly similar candidates $S$ and $R$. For details see \cite{Brubach}, which includes a justification for why this notion of candidate similarity may be more useful than other measures.

As with the first approach, we note that when $S$ is strongly similar to $A$, this looks like the ``classic'' vote-splitting effect in Table~\ref{votesplitting}. However, when $S$ is strongly similar to $R$, the spoiler candidate appears to be somehow helping a similar candidate $R$ win rather than splitting the vote with $A$.

\begin{table}[]
    \centering
    \begin{tabular}{r||c||c|c||c|c}
        Method & \multicolumn{1}{p{10 mm}||}{Spoiler events} & \multicolumn{1}{p{17 mm}|}{$S$ voters prefer $A$ \%} & \multicolumn{1}{p{17 mm}||}{$S$ voters prefer $R$ \%} & \multicolumn{1}{p{12.5 mm}|}{$S$-$A$ str. sim. \%} & \multicolumn{1}{p{12.5 mm}}{$S$-$R$ str. sim. \%} \\
        \hline \hline 
        
        Bloc & 116 & 51 & 49 & 3 & 15 \\
        $k$-Borda OM & 224 & 6 & 94 & 0 & 33 \\
        $k$-Borda PM & 68 & 74 & 26 & 24 & 0 \\
        SRCV & 27 & 100 & 0 & 4 & 0 \\
        MM Cond. Comm. & 4 & 100 & 0 & 0 & 0 \\
        SNTV & 159 & 97 & 3 & 41 & 0 \\
        Top $k$ IRV & 6 & 83 & 17 & 0 & 0 \\
        Cham-Cour OM & 171 & 23 & 77 & 1 & 11 \\
        Cham-Cour PM & 117 & 39 & 61 & 0 & 7 \\
        Scottish STV & 66 & 61 & 39 & 2 & 3 \\
        Meek STV & 24 & 92 & 8 & 0 & 0 \\
        EAR OM & 383 & 29 & 71 & 1 & 13 \\
        EAR PM & 238 & 29 & 71 & 1 & 6 
    \end{tabular}
\caption{The first column lists the number of times removing a single spoiler candidate caused one swap in the winning set for each method. The second and third columns show how often the voters ranking the spoiler first preferred the alternate winner versus the real winner when a spoiler event occurred. The two rightmost columns give the percentage of spoiler events in which the spoiler was strongly similar to the alternate or real winner, respectively.}
    \label{tab:new_splitting_table}
\end{table}

This analysis of vote-splitting is summarized in Table~\ref{tab:new_splitting_table}.
For each method, we refer to a \emph{spoiler event} as a triple of distinct candidates $(R,A,S)$. The entries under ``$S$ voters prefer $A$ \%'' give the percentage of spoiler events in which voters ranking the spoiler $S$ first prefer the alternative winner $A$. The entries under ``$S$ voters prefer $R$ \%'' capture the opposite scenario. The columns ``$S$-$A$ str. sim. \%'' and ``$S$-$R$ str. sim. \%'' give the percentage of spoiler events in which $S$ is strongly similar to $A$ and $R$, respectively. 
Spoiler events have a greater potential to be caused by vote-splitting in methods with higher percentages of $S$ voters preferring $A$ or higher percentages of $S$ being strongly similar to $A$. When these percentages are greater than their counterparts, a candidate is more likely to be hurt than helped by the presence of a similar candidate. 
Conversely, methods with higher percentages of $S$ voters preferring $R$ or higher percentages of $S$ being strongly similar to $R$ can be seen as ``friendly'' to similar candidates running and less susceptible to vote-splitting as a cause of spoiler events. In these methods, candidates would like to have weaker versions of themselves run, as this can help boost the candidate to victory.

As with previous spoiler results, we do not see any strong correlation between a method's categorization as proportional or majoritarian and the method's tendency toward vote-splitting.  
Proportional methods based on Cham-Cour tend to produce spoiler results which are not best understood as vote-splitting while most of the spoilers found under Meek STV appear to involve vote-splitting. 

We note that while SRCV and Meek STV have high percentages of $S$ votes preferring $A$, they have relatively few spoiler events overall and low percentages of events in which $S$ and $A$ are actually similar candidates. Rather, under these methods, $S$ is merely more similar to $A$ than it is to $R$. 
$k$-Borda PM and SNTV are the only methods that display a substantial percentage of events in which $S$ and $A$ are strongly similar candidates, suggesting that the candidates ``split'' the vote, and causing $A$ to lose. In Section~\ref{subsec:addclones}, we will see that $k$-Borda PM and SNTV are also the two methods in which adding a clone of a winner is most likely to cause that winner to lose.

For methods with high percentages of $S$ voters preferring $R$ or high percentages of strongly similar $S$ and $R$, vote-splitting is not a major factor, and we must look for new causes of the spoiler effect. 
While $k$-Borda OM, Cham-Cour OM, and EAR OM have large numbers of spoiler events, the spoiler candidates are helping similar candidates rather than splitting the vote. This could be driven partly by the OM model. Under that model, a ballot ranking only $A$ followed by $S$ will reduce the relative Borda score of other candidates more than a ballot ranking $A$ alone.

\subsection{A closer look at STV and vote-splitting}
\label{subsec:stvcloserlook}

\begin{figure*}[tp]
    \centering
    \includegraphics[width=0.86\linewidth]{"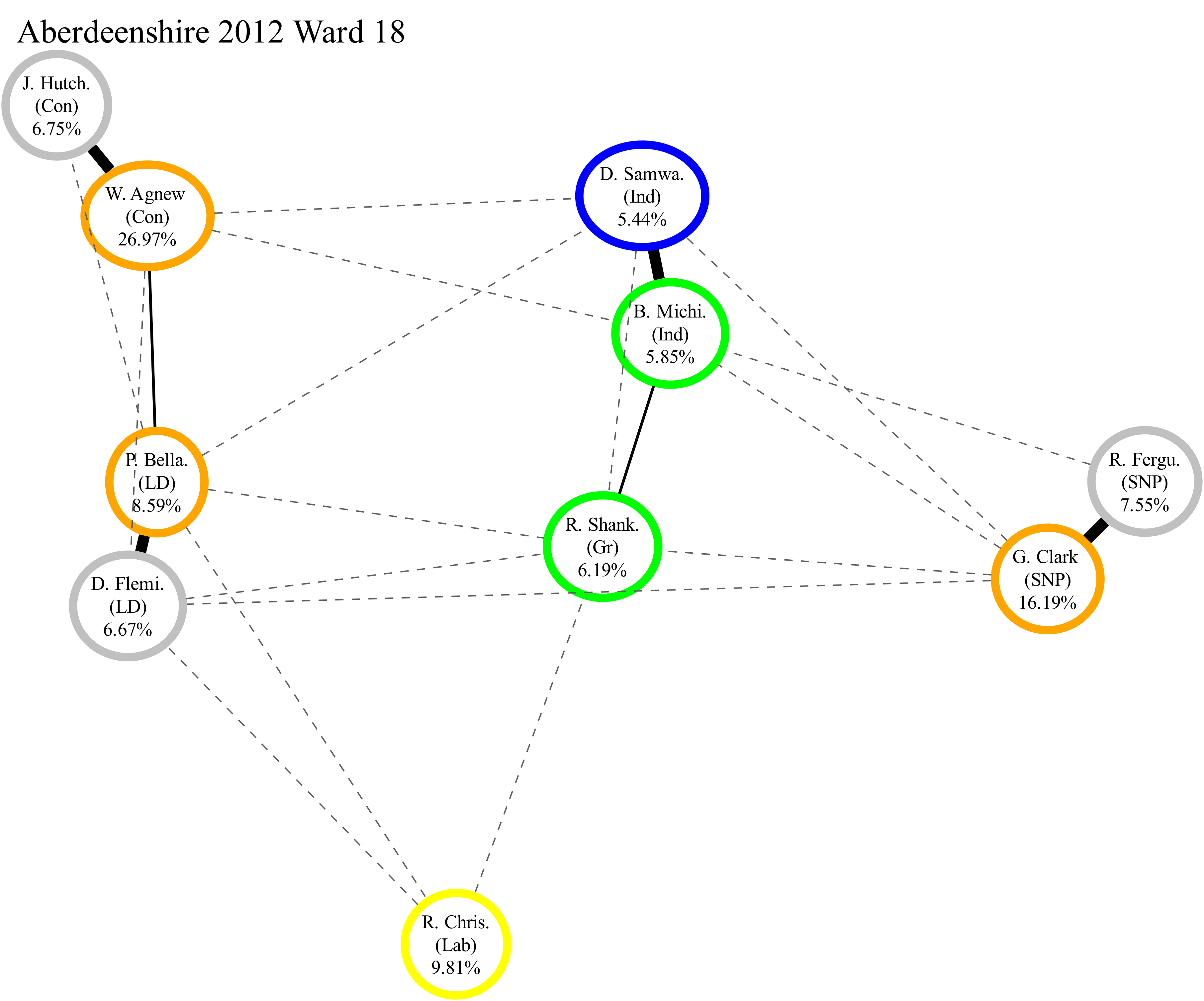"}

    \medskip
    \includegraphics[width=0.72\linewidth]{"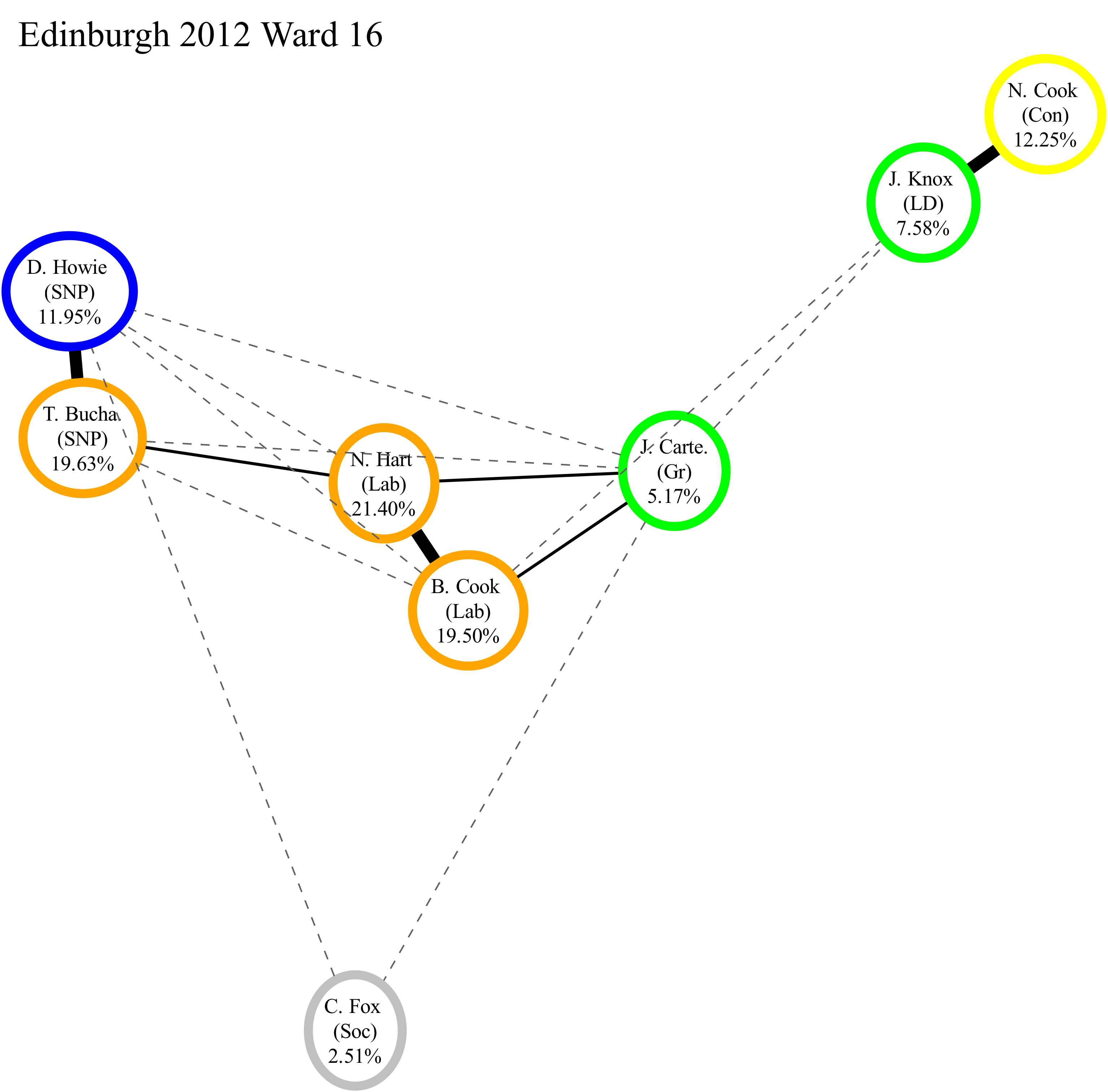"}
    \quad
    \includegraphics[width=0.2\linewidth]{"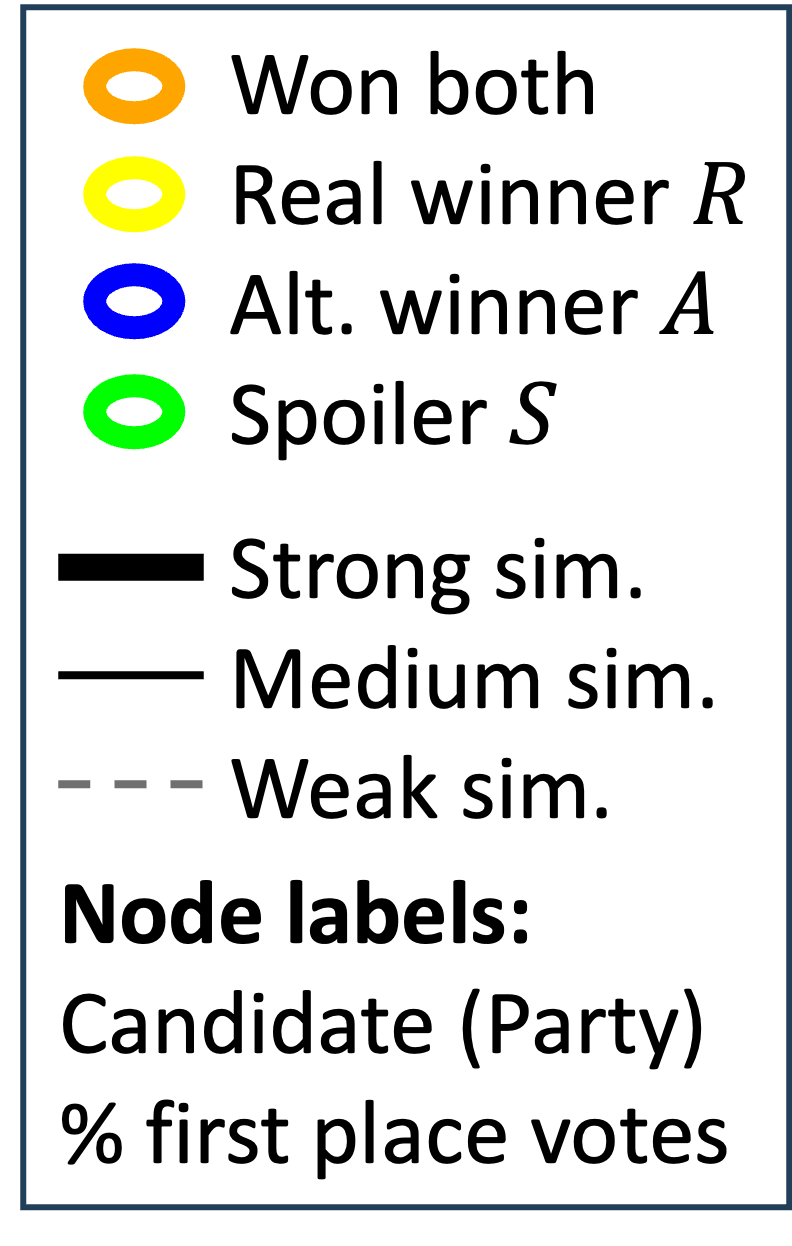"}
    \caption{Mutual preference graph visualizations of two Scottish STV elections which each had two distinct spoilers. The yellow (real winner) and orange candidates won in the real world election. However, removing one of the green candidates (spoilers) causes the blue candidate (alternate winner) to replace the yellow candidate in the winning set.}
    \label{fig:spoilergraphsbody}
\end{figure*}

To better understand the relationship between  spoiler events and vote-splitting under Scottish STV, we look more closely at the circumstances under which the change in winner occurs. Several  of the Scottish STV events  that appear to be due to vote-splitting between $S$ and $A$ also involve $A$ being eliminated before $S$. This suggests that vote-splitting spoilers in Scottish STV are driven by relatively popular, competitive spoiler candidates as opposed to less popular, ``fringe'' candidates. On the other hand, when the $S$ voters prefer $R$ in Scottish STV spoilers, we find that $A$ is often strongly similar to another winning candidate. In this case, $A$ may receive more spillover votes from that winning candidate when the quota is lowered due to removing voters who only ranked $S$. This also offers an explanation for why Meek STV, which involves lowering the quota while counting votes, has many fewer spoiler events in which $S$ voters prefer $R$ compared to Scottish STV. Further, these observations align with the fact that Scottish STV spoilers have relatively high plurality strength and Meek STV spoilers have even higher plurality strength which generally exceeds the plurality strength of the alternate winners as seen in Figure~\ref{figure:spoiler_relative_strength}.

Figure~\ref{fig:spoilergraphsbody} provides visualizations of the relationships between candidates in two Scottish STV elections. Each node represents a candidate, with the percentage indicating the proportion of ballots ranking that candidate first. The graphs position candidates in two dimensions according to their Mutual Preference Distance, a measure of candidate similarity based on the preferences of voters who rank each candidate first (see \cite{Brubach} for details). In addition to strong similarity as defined above, the graphs show medium and weak similarity. Specifically, thick solid edges indicate strong similarity (Mutual Preference Distance equal to 1), thin solid edges indicate medium similarity (distance greater than 1 and at most 2), and dashed edges indicate weak similarity (distance greater than 2 and at most half the number of candidates). Candidates without a connecting edge are considered dissimilar. Because Mutual Preference Distances cannot generally be represented exactly in two dimensions, the edge types provide a more reliable indication of similarity than the apparent distances between nodes.

Each election in Figure~\ref{fig:spoilergraphsbody} has two distinct spoiler candidates, highlighted in green, whose individual removal changes the winning committee. The top graph illustrates an election in which both spoilers are relatively close to the alternative winner, who is eliminated before either spoiler under Scottish STV. The 2012 election in Aberdeenshire Ward 18 is the only Scottish STV election in our dataset where $S$ is strongly similar to $A$. The bottom graph illustrates a different phenomenon: one of the spoilers is strongly similar to the real winner $R$. In Edinburgh 2012 Ward 16, $A$ is closer to a cluster of other winners than $R$, potentially allowing $A$ to receive a larger share of surplus transfers when removing a spoiler lowers the quota. Additional mutual preference graphs for the 2022 Edinburgh Ward 5 election and the 2012 Highland Ward 7 election, discussed in Example~\ref{edinburgh} and Table~\ref{table:highland_2012_ward7} (Section~\ref{section:PR}), respectively, appear in \ref{appendix_section:graphs}.

\subsection{Adding clones}
\label{subsec:addclones}

In this section we consider the question of spoilers from the opposite direction: if a candidate enters a race. To investigate this possibility, we investigate how the winner set changes when a clone of an existing candidate is added to the preference profile.  We change the profile by  selecting an existing candidate (the \emph{cloned candidate}) and insert a new candidate (the \emph{clone})  just below them or just above them in each ballot in which the existing candidate appears. To ensure that the clone is tied or weaker than the cloned candidate, we assume that the clone is ranked just below the cloned candidate except in a specific fraction of cases (more details below). 

The addition of a clone to the ballot data can change the winner set in multiple ways. For example, it is possible that the cloned candidate loses  in the original election but wins with the addition of the clone, or vice versa. It is possible the clone itself joins the winner set.  There may also be changes to the winner set that do not involve either the cloned candidate or the clone. 
Not all of these outcomes are of equal concern. Arguably, the two most serious outcomes are:

\begin{enumerate}
\item a losing candidate is cloned and both the losing candidate and the clone win in the new election. If a method were susceptible to this kind of outcome, parties would be incentivized to run many similar candidates.
\item a winning candidate is cloned and loses in the resulting profile, a ``vote-splitting'' kind of outcome. If a method were susceptible to this kind of outcome,  parties would be incentivized to encourage clones of opposing party candidates to run.
\end{enumerate}

Table  \ref{clone_cases} lists all the possible outcomes we consider when adding a clone. The letter before the arrow indicates if the cloned candidate was a winner ($W$) or a loser ($L$). After the arrow, the first entry denotes whether the cloned candidate wins ($W$) or loses ($L$) a seat, and the second entry denotes the outcome for the clone. For instance, $L \rightarrow W,L$ denotes that a losing candidate was cloned, after which the cloned candidate wins and the clone loses. Note that since the clone is never stronger than the cloned candidate, it is not possible to have the combination $L \rightarrow L,W$ or $W \rightarrow L,W$. We also consider the case where a clone is added, which does not affect the cloned candidate, but does cause some other change in the winner set. We denote these outcomes $L \rightarrow L,L + other$ and $W \rightarrow W,L + other$.

 \cite{Woodall} formalizes some of these possibilities in his \emph{clone-no-help} and \emph{clone-no-harm} axioms. Clone-no-help requires that replacing a candidate by a set of clones does not cause a candidate outside the clone set to change from losing to winning, while clone-no-harm requires that it does not cause a candidate outside the clone set to change from winning to losing. Several of the outcomes in our classification correspond directly to violations of these properties. In particular, both $L\rightarrow L,L+\ other$ and
$W\rightarrow W,L+\ other$ violate both properties, since a change among candidates outside the clone pair necessarily causes one previously losing candidate to win and one previously winning candidate to lose.

Winner set changes that do not involve the cloned candidate or the clone are less  concerning in our view. Although such outcomes violate Woodall's  properties and may have meaningful impacts on representation and policy, their consequences are less predictable than the more serious clone types, making it less clear how a candidate or party might exploit them strategically. Thus, Table \ref{clone_cases} lists the outcomes in order of significance, with outcomes at the top of the table being more normatively concerning in our view than those at the bottom.

\begin{table}
\begin{tabular}{@{} l @{\hskip 1pt} l @{}}
Type: & Description\\
\hline 
   $L \rightarrow W, W$:& cloning a losing candidate causes both cloned candidate and clone to win \\
$L \rightarrow W, L$: &  cloning a losing candidate causes the cloned  candidate to win \\
$W\rightarrow L, L$: &   cloning a winning candidate causes the cloned candidate  and clone to  lose\\
$L \rightarrow L, L+ other$: & cloning a losing candidate causes other winners to change \\
$W\rightarrow W, L+ other$:&  cloning a winning candidate causes  other winners to change \\
$W \rightarrow W, W$: &  cloning a winning candidate causes both cloned candidate and clone to win \\
\end{tabular}
\caption{Types of clone behavior arranged in descending order of significance}\label{clone_cases}
\end{table}

To measure the impact of the ``strength'' of the clone on the election outcome, we consider three scenarios: (i)  the clone lies just below the cloned candidate in every ballot; (ii) the clone lies just above the  cloned candidate in 10\% of the ballots and just below the cloned candidate in the remaining ballots; (iii)  the clone lies just above the cloned candidate in 50\% of the ballots and just below the cloned candidate in the remaining ballots. Specifically, for (ii) and (iii), we consider every kind of ballot that was cast, count how many copies of that ballot were cast, and then place the clone above the cloned candidate in 10 or 50\% of those ballots, rounded down, and place the clone below the cloned candidate in the remaining ballots. In all three cases, the clone is tied or weaker than the cloned candidate. 

Table \ref{clone_results} shows the number of elections in which the addition of a clone changed the winning committee. The table breaks down the results based on the kind of clone outcome, and by the strength of the clone added. For example, for bloc voting there are 5 elections where there exists a losing candidate $C$ such that adding a ``strong clone'' of $C$ (a clone who is just above $C$ on almost  50\% of ballots where $C$ appears) causes both $C$ and their clone to join the winning committee. 

One election can have multiple different clones that can change the outcome. Table \ref{clone_results_clone_count} in  \ref{sec:clone_appendix} reports the total number of individual clones found to change the winning committee.

\begin{table}
\centering
{\small 
\begin{tabular}{c||ccc||ccc}
&\multicolumn{2}{c}{$L \rightarrow W, W$ }& & \multicolumn{2}{c}{$L \rightarrow W, L$} \\
\hline \hline
Method & 0\% & 10\% & 50\%  & 0\% & 10\% & 50\% \\
\hline \hline
Bloc  & 1 & 1 & 5 & 302 & 285 &136\\
$k$-Borda OM & 0 & 0 & 43 & 770 &758 &479\\
$k$-Borda PM & 0 & 0 & 0 & 223 & 181 & 4\\
SRCV & 0 & 0 & 0 & 0 & 0 & 0\\
MM Cond Committee & 0 & 0 & 0 &2 & 2 & 2\\
SNTV& 0 & 0 & 0 & 0 & 0 & 0\\
Top $k$ IRV & 0 & 0 & 0 &3 & 3 & 3\\
Cham-Cour OM & 0 & 0 & 0 & 864 & 845 &647\\
Cham-Cour PM & 0 & 0 & 0 & 644	 & 604 & 285\\
Scottish STV& 0 & 0 & 0 & 0 & 0 & 0\\
Meek STV & 0 & 0 & 0 &0 & 0 & 0\\
EAR OM & 0 & 0 & 0 &776 & 766 & 669\\
EAR PM & 0 & 0 & 0 &580 & 562 & 408\\
\hline
\end{tabular}

\vspace{0.1 in}

\begin{tabular}{c||ccc||ccc}
\hline
&\multicolumn{2}{c}{$W \rightarrow L, L$}& & \multicolumn{2}{c}{$L \rightarrow L, L+other$} \\
\hline
Method & 0\% & 10\% & 50\%  & 0\% & 10\% & 50\% \\
\hline
Bloc  & 0 & 2 & 89 & 120 & 122 & 127\\
$k$-Borda OM & 0 & 0 & 0 & 80 & 81 & 102\\
$k$-Borda PM & 0 & 0 & 104 & 30 & 31 & 35\\
SRCV & 0 & 0 & 0 & 0 & 0 & 0\\
MM Cond Committee & 0 & 0 & 0 &0 & 0 & 0\\
SNTV& 0 & 139 & 793 & 0 & 0 & 0\\
Top $k$ IRV & 0 & 0 & 0 &0 & 0 & 0\\
Cham-Cour OM & 0 & 0 & 0 & 46 & 47 & 56\\
Cham-Cour PM & 0 & 0 & 0 &46 & 47 & 48 \\
Scottish STV& 0 & 0 & 0 & 0 & 0 & 0\\
Meek STV & 0 & 0 & 0 &0 & 0 & 0\\
EAR OM & 0 & 0 & 10 &97 & 101 & 123\\
EAR PM & 0 & 0 & 2 &61 & 62 & 68\\
\hline
\end{tabular}

\vspace{0.1 in}

\begin{tabular}{c||ccc||ccc}
\hline
&\multicolumn{2}{c}{$W \rightarrow W, L + other$}& & \multicolumn{2}{c}{$W \rightarrow W, W$} \\
\hline
Method & 0\% & 10\% & 50\%  & 0\% & 10\% & 50\% \\
\hline
Bloc  &62 & 59 & 23 &   937 & 946 &990\\
$k$-Borda OM & 69	 & 55	 & 3     & 895 & 939 & 998\\
$k$-Borda PM & 54 & 44 & 22    &841 & 868 & 970 \\
SRCV & 0 & 0 & 0 & 999	& 999 & 999\\
MM Cond Committee & 0 & 0 & 0 &977 & 977 & 988\\
SNTV& 0 & 0 & 0 & 0 & 0 & 301\\
Top $k$ IRV & 0 & 0 & 0 &0 & 0 & 268\\
Cham-Cour OM & 0 & 39 & 202     & 0 & 0 & 132\\
Cham-Cour PM & 35 & 44 & 126    &0 & 0 & 6 \\
Scottish STV& 12 & 16& 36 &  191 & 191 & 259 \\
Meek STV & 3 &3 & 11   &251 & 251 &	320 \\
EAR OM & 176 & 191 & 167 &409 & 408 & 656\\
EAR PM & 74 & 85 & 95 &427 & 434 & 538\\
\hline
\end{tabular}
}
\caption{Number of Elections that are affected by the addition of a clone.}\label{clone_results}

\end{table}

Not all clone outcomes are possible under each voting method. For instance, under SNTV, if a losing candidate is cloned, it is not possible for the clone to become a winner: since the  losing candidate does not have one of the top $k$ plurality scores, the same will be true of its clone, regardless of the proportion of ballots in which the clone is inserted above the candidate. Thus all the entries in Table \ref{clone_results}, corresponding to $L \rightarrow W, W$ and $L \rightarrow W, L$ under SNTV must be 0 regardless of the preference profile. There may be additional entries for which this is true.
 
Table \ref{clone_results} illustrates several clone-related features of the various voting rules. There are almost no instances of the $L \rightarrow W,W$ outcome for any method, but there are many elections in which  Borda OM, Cham-Cour, and EAR demonstrate $L \rightarrow W,L$ outcomes. This reinforces the findings of Table \ref{tab:new_splitting_table} in which these methods appear to be relatively clone-friendly:  parties are incentivized to run weaker candidates to help their stronger, more viable candidates win seats. On the other hand, SNTV is most susceptible to $W \rightarrow L,L$ outcomes, followed by Borda PM, again agreeing with the findings in Table \ref{tab:new_splitting_table} that suggest these methods are susceptible to classic vote-splitting.

Notably, both versions of STV are quite resistant to clone-related outcomes, giving large  numbers only for the $W\rightarrow W, W$ type of outcome. It is unsurprising that STV performs well under these measures, given that in the single-winner case STV is independent of clones in Tideman's sense \cite{Ti87}. Some of the \mbox{$W \rightarrow W, W$} type outcomes can perhaps be explained by the fact that parties never run $k$ candidates, and thus it is plausible that a party could often earn an extra seat if it ran another candidate. For example, in 32 of the 999 elections a party runs a single candidate and that candidate earns an initial first-place vote total in excess of 1.75$\times$quota; similarly, in 25 elections a party runs two candidates and those two candidates combined earn a first-place vote tally in excess of 2.5$\times$quota. In these circumstances it is neither surprising nor concerning that clone of a winner would join the winner set. Furthermore, while such $W \rightarrow W, W$ outcomes are violations of Woodall's clone-no-harm criterion, Woodall observes that this criterion is incompatible with proportionality, and proportionality seems like a more important criterion.

While high numbers of $W\rightarrow W, W$ outcomes may not be immediately concerning, they can help elucidate the priorities of a voting method. For example, unless there are tied outcomes then a clone of the candidate who wins the first seat under SRCV is guaranteed to win the second seat; i.e., in any multiwinner election the clone of the first seat is guaranteed to join the winning committee. This is why SRCV achieves the maximum value of 999 in all three columns, and helps show that SRCV is firmly majoritarian, making it inappropriate for specific kinds of elections.

Our clone and vote-splitting results paint a nuanced picture of the relationship between spoilers and vote-splitting under STV. When an election exhibits a spoiler effect, the spoiler is somewhat more likely to resemble the candidate who would have won in its absence, consistent with the traditional vote-splitting mechanism. Nevertheless, artificially introducing exact clones almost never produces the strategically problematic forms of spoiler behavior we might fear. Thus, STV is not entirely immune to vote-splitting effects, but STV is highly resistant to spoiler effects caused by the presence of a clone.

\section{The Spoiler Effect and Proportional Representation in Scottish STV}\label{section:PR}

The existence of a spoiler effect tells us that the composition of the winning committee is unstable with respect to the removal of a losing candidate. It does not, by itself, tell us that proportional representation (PR) has been violated in some way. 

For example, if every spoiler instance involved the swapping of candidates from the same party in the winner set then the 66 Scottish STV spoiler instances would presumably not represent some sort of PR failure. In this section we focus on the spoiler effect instances for Scottish STV, since this method is actually used for the elections in our dataset. Because there are only 66 spoiler instances, we thoroughly examine each instance individually to see how they occur and what the PR ramifications are. 

To this end, this section answers three questions for Scottish STV:

\begin{enumerate}
    \item When a spoiler effect occurs, how substantively large is the change in the composition of the winning committee?
    \item When a cross-party candidate swap occurs, under what electoral conditions does it occur?
    \item Does the original winning committee or the winning committee with the spoiler removed violate standard PR axioms?
\end{enumerate}

For (1), in five of the spoiler instances $R$ and $A$ belong to the same party; while the spoiler effect may be concerning to the individual candidate whose win was ``spoiled'' or to the voters who ranked the would-be winner in first, such instances seem unimportant from a PR perspective. For the other 61 instances, $R$ and $A$ are often quite dissimilar, as shown in Table \ref{tab:new_splitting_table}. In some instances the presence of the spoiler causes a party to earn one seat instead of two, resulting in a 50\% decrease in party representation. Such changes to the winning committee are substantive.

For (2), the majority of the remaining 61 instances occur in elections that are highly competitive. Thus, it is unsurprising that any small perturbation to the election, including the removal of a candidate, can cause the election to unfold in a different order, eventually changing the winning committee. We see this dynamic in the election shown in the top of Table \ref{e-duns-table}, where in the final round, Green Party candidate Bandel defeats the Liberal Democrat Wood by a  vote margin of less than a percentage point. When we remove a candidate like Herring, it is perhaps unsurprising that Wood can replace Bandel in the winning committee, even though Bandel and Wood are not ideologically similar.

The most extreme such competitive outcome is shown in Table \ref{table:highland_2012_ward7}. In the penultimate round Rattray defeats Fletcher by approximately one hundredth of a vote; if we remove Rous, Fletcher wins. The competition between Fletcher and Rattray is so close that even a very small perturbation to the election can determine which candidate receives the final seat. The presence of Rous provides  such a perturbation, causing the count to fall in Rattray's favor. See Figure \ref{fig:spoilergraphsappendix1} in  \ref{appendix_section:graphs} for the election's mutual preference graph.

Excluding the five instances in which $R$ and $A$ belong to the same party, 27  spoiler instances have the following structure: $R$ and $A$ are the two candidates remaining in the final round of the election with one seat remaining, and $R$'s margin of victory over $A$ is less than one percentage point. Expanding the margin to two percentage points increases this number to 33. Thus, while many of the spoiler effect instances lead to a swap of potentially dissimilar candidates, the electoral competitiveness could explain the outcome. Informally, in the election in Table \ref{table:highland_2012_ward7}, it is not clear if there is a large enough bloc of voters within the ``Liberal Democrat'' camp to deserve a seat, or if there is a large enough bloc of SNP voters to warrant two seats. Because the election is knife-edge, either outcome seems reasonable.

\begin{table}
{\small
\begin{tabular}{l|c|c|c|c|c|c|c}
\multicolumn{8}{c}{Quota $= 651$, $k=4$}\\
\hline
Candidate & \multicolumn{7}{c}{Votes by Round} \\
\hline
Wilson (Ind) & \textbf{768} &  &  &  &  &  &  \\
Rous (Con) & 65 & 68.50 & 0 & 0 & 0 & 0 & 0 \\
MacInnes (Lab) & 324 & 332.23 & 336.53 & 367.45 & 367.86 & 0 & 0 \\
Finlayson (Ind) & 389 & 429.37 & 452.28 & 519.03 & 519.80 & 621.30 & \textbf{777.14} \\
Rattray (LD) & 361 & 378.67 & 394.59 & 481.41 & 483.38 & 547.96 & \textbf{630.98} \\
Fletcher (SNP) & 464 & 469.48 & 470.64 & 489.25 & 494.96 & 547.95 & 0 \\
McCaffery (Ind) & 299 & 307.07 & 313.23 & 0 & 0 & 0 & 0 \\
Smith (SNP) & 584 & 604.72 & 606.87 & \textbf{660.78} &  &  &  \\
\hline
\end{tabular}}

\caption{The votes in each round for candidates in the 2012 election in Ward 7 of the Highland council area. Fletcher loses in the penultimate round by approximately one-hundredth of a vote.}
\label{table:highland_2012_ward7}

\end{table}

Electoral competitiveness provides one partial explanation of why a spoiler effect might arise. Other features of the STV tabulation procedure may also provide insight. For example, when we remove a losing candidate we also remove voters who rank only that candidate, resulting in a lower quota in the modified profile. It is possible that this change in the quota, rather than changes in the distribution of preferences among the remaining candidates, is primarily responsible for some spoiler effects.  To examine this possibility, for each spoiler instance $(R, A, S)$ we remove $S$ from all ballots and calculate the Scottish STV winning committee for $(\mathcal{P}^{-S}, \mathcal{C}^{-S}, k)$ but keep the quota from $(\mathcal{P}, \mathcal{C},k)$. Of the original 66 spoiler instances, 15 no longer demonstrate a spoiler effect when we keep the quota fixed. In these cases, the spoiler effect may be better understood as a consequence of STV's sensitivity to the exact choice of quota than as something like vote splitting or clone sensitivity.

Ultimately, across the 66 spoiler instances, we could not find a single unifying mechanistic explanation for why STV sometimes exhibits the spoiler effect. STV tabulation is a complicated process, and thus it is unsurprising that spoiler effects might arise in a variety of ways.

Competitiveness and quota changes provide only indirect evidence about the PR consequences of a spoiler effect. A close contest between $R$ and $A$ does not imply that both resulting committees (the actual winner set and the winner set with the spoiler removed) provide adequate representation to the electorate. For example, it is possible that one committee excludes a candidate supported by a sufficiently large and cohesive group of voters, while the other does not. To examine this possibility, addressing question (3), we turn to an axiomatic analysis of the two winning committees. Rather than focusing on the margin separating $R$ and $A$, we ask whether either committee fails standard proportionality axioms.

For each spoiler instance, we consider both the original winning committee $\mathcal{W}$ and the winning committee obtained after removing the spoiler $\mathcal{W}^{-S}$. Moreover, because removing the spoiler changes the preference profile itself, we evaluate both committees on both $\mathcal{P}$ and $\mathcal{P}^{-S}$. We consider three standard proportionality axioms: proportionality for solid coalitions (PSC) \cite{D}, local stability \cite{AEFLS},  and rank-EJR+ \cite{BP}, using both the Hare and integer Droop quotas. For each axiom a given winning committee  can either violate or ``pass'' the axiom for a given preference profile, and thus we can test whether the original winning committee violates any of these axioms on the modified profile with the spoiler removed, for example. Precise definitions of the axioms can be found in  \ref{appendix:proportionality}. Informally, each axiom stipulates what kind of representational outcomes should occur for large voter blocs who vote cohesively. The three axioms are logically independent, and Scottish STV satisfies PSC but not the others.

For each Scottish STV spoiler instance $(A, S, R)$,  we checked if $\mathcal{W}$ or $\mathcal{W}^{-S}$  violated any axiom on $\mathcal{P}$ or $\mathcal{P}^{-S}$. Since Scottish STV satisfies both Droop and Hare PSC,  $\mathcal{W}$ cannot violate PSC for the original profile and $\mathcal{W}^{-S}$ cannot violate PSC on the modified profile.

In every spoiler instance, both winning committees satisfy all six proportionality axioms on both profiles. In particular, $\mathcal{W}$ satisfies all axioms on $\mathcal{P}^{-S}$ and $\mathcal{W}^{-S}$ satisfies all axioms on $\mathcal{P}$. Thus, we find no spoiler instance in which a change in the winning committee comes at the expense of PR according to any of these measures. This suggests that the observed Scottish STV spoiler effects are less concerning from a PR perspective than the change in winner set might initially suggest.

We note these results are not simply a consequence of the proportionality conditions being weak on these profiles. With the exception of PSC using the Hare quota, the axioms are quite restrictive. See  \ref{appendix:proportionality} for further analysis. 

\section{Worst Case Analysis}\label{section:worst_case}

In most of our real-world elections, if a voting method demonstrates a spoiler effect then there is only one spoiler candidate and the removal of this candidate changes the winner set by only one candidate. In theory, when we have complete control over the preference profile, the effect of removing a spoiler candidate can be much more extreme. In this section, we investigate how extreme spoiler effects can be under various methods. 

There are various ways to define a ``worst-case'' spoiler outcome. One possibility is profiles in which different spoilers can result in disjoint winner sets.  In this discussion, we focus on what we term \emph{extreme spoiler candidates}, which are candidates whose removal causes the entire winner set to change. 

\begin{definition}
For a given voting rule $f$ and election $(\mathcal{C},\mathcal{P}, k)$, an \textbf{extreme spoiler} is a losing candidate $S$ such that $f(\mathcal{C}, \mathcal{P},k) \cap f(\mathcal{C}^{-S},\mathcal{P}^{-S},k)=\emptyset$, assuming there are no ties. That is, assuming $\mathcal{W}$ and $\mathcal{W}^{-S}$ are unique, $S$ is an extreme spoiler if $\mathcal{W}\cap \mathcal{W}^{-S} =\emptyset$. \end{definition}

In this paper, the worst-case scenario is when there are $k+1$ losers, each of whom is an extreme spoiler. Thus, the removal of any one loser changes the status of every single other candidate in the election. 

STV is quite susceptible to spoilers, with the caveat that the first candidate to be eliminated has limited impact on the election and cannot be an extreme spoiler, as stated in the following proposition which is proven in  \ref{sec:worst_case_appendix}:

\begin{proposition} There exists a profile in which under STV, SRCV, or Top-$k$ IRV, every non-winning candidate except the first candidate eliminated is an extreme spoiler.
\end{proposition}

An informal explanation of the proof is that in a constructed profile the original winners can be given only a slight advantage over the losing candidates, so that removing any one of the latter changes the order of elimination and ultimately produces an entirely different winning committee. The exception is the first candidate eliminated, whose removal cannot produce such a complete change.

For many of the other voting methods, it is possible to construct a preference profile in which every losing candidate is an extreme spoiler.  Perhaps the most resistant to extreme spoilers is the Minimax Condorcet Committee method which can have extreme spoilers, but there is a limit to their effect: in order for a loser to be an extreme spoiler, there must be other losers in the election that do not become winners when the spoiler is removed. 

Full proofs of the worst-case scenarios are provided in ~\ref{sec:worst_case_appendix}. The proofs require extremely unrealistic preference profiles, partially explaining why we do not see such extreme spoiler effects in the real elections. Thus, in theory spoiler effects can be quite extreme for the methods we study, but in practice we see nothing close to such worst-case outcomes.

\section{Conclusion}\label{section:conclusion}
We can draw several primary conclusions. Firstly, overall susceptibility to spoilers appears not to be closely tied to the proportionality  of the voting method.
As illustrated in Table \ref{tab:results_table}, majoritarian methods such as $k$-Borda OM and Bloc fare poorly as do proportional methods such as EAR OM, EAR PM and Cham-Cour OM.  This lack of correspondence also occurs in Table \ref{clone_results} where these same methods exhibit higher incidences of sensitivity to different kinds of cloned candidates.

In contrast, both variants of STV exhibit fewer instances of spoilers than any of the other methods except for top-$k$ IRV and Cond Committee, which were specifically included because of their resistance to spoilers.  Both Scottish STV and Meek STV also performed well based on  other measures: there were few instances of multiple spoilers and  few instances of weak spoilers. The STV winner sets also appear relatively robust to the addition of clones. The most frequent change to the winner set occurs  when a winning candidate is cloned, resulting in both the cloned candidate and clone winning. This is arguably of minimal concern as it indicates that the support for the cloned candidate was strong enough to ``warrant'' another similar winner. 

This fact, as well as the relative strength of the spoiler candidates under STV, indicates that when a change of winner set occurs, it is not the result of a fringe candidate entering or leaving the race. Nor is it often the result of traditional vote splitting. As illustrated in Table  \ref{tab:new_splitting_table}, in a significant fraction of spoiler events,  voters who top-ranked the spoiler candidate preferred the real winner to the alternate, throwing into question the designation of  ``spoiler.''

Overall, the analysis suggests that spoiler effects are not overly concerning under Scottish STV and Meek STV. While the elimination of a candidate can affect the winner set, this happens generally when the election is close. As discussed in Sections \ref{section:results_from_scottish_data} and \ref{section:PR}, the elimination of a candidate can affect the winner set in multiple distinct ways related to changes in quota, vote transfer rates, and the disappearance of bullet votes for the spoiler candidate. It is not surprising, therefore, that changes to the candidate set will sometimes alter the result.

Most importantly, from a proportional representation standpoint, the elimination of the spoiler candidate did not affect the behavior of either the real or alternate winner set with respect to  three of the standard proportionality axioms. This suggests that spoiler susceptibility under Scottish STV does not represent some kind of failure of proportional representation.\\

\bibliographystyle{plainnat}

\bibliography{sn-bibliography}

\appendix
\renewcommand{\thesection}{Appendix \Alph{section}:}
\section{Multiwinner voting rules}\label{voting_rules_app}

We define our voting methods more fully here. Recall that because ties are rare in our analysis, we do not discuss how to break ties as part of our definitions.

\begin{itemize}

\item \textbf{Bloc Voting (Bloc)}: the winning committee consists of the $k$ candidates receiving the highest $k$-approval score, which is the number of voters who rank the candidate among their top $k$ candidates.

\item $\mathbf{k}$\textbf{-Borda}: Candidates are assigned points based on their position on individual ballots. After summing, the $k$ candidates with the highest total Borda score form the winning committee. For a given ballot, a first-place ranking is worth $m-1$ points, a second-place ranking is worth $m-2$ points, and so on.

\item \textbf{Sequential Ranked-choice voting (SRCV)}: (SRCV)  works by sequentially filling seats using the single-winner implementation of STV. The first seat is given to the winner of the election when using STV but assuming $k=1$. That is, we eliminate the candidate with the fewest first-place votes and transfer their ballots as if using STV, until a candidate earns a majority of the first-place votes. That candidate is then given the first seat, and is eliminated from the ballot data. We repeat the process, again using STV assuming $k=1$, on the modified ballot data (where the first winner has been removed) to assign the second seat; the resulting winner is given the second seat and eliminated from the ballot data. This process continues until all $k$ seats are filled. 
SRCV is used rarely and is little studied;  \cite{MMLS24} compare SRCV to STV, finding that its behavior is closer to  majoritarian-based methods rather than  proportional methods such as STV. We include SRCV because it has been used in real-world elections, most notably in city council elections for several small cities in the state of Utah in the US.
\item \textbf{Single non-transferable vote (SNTV)}: the winning committee consists of  the $k$ candidates with the largest number of first-place votes. 

\item \textbf{Chamberlin-Courant (Cham-Cour)}: There are a number of variants this rule;  we use the original method proposed in \cite{Cham-Cour}. Let $\mathcal X \subset \mathcal C$ be a subset of candidates of size $k$. For each $C_i \in \mathcal X$, let $V_{i}(\mathcal X)$ denote the set of voters for whom candidate $C_i$ is the most preferred candidate in the committee $\mathcal X$.
If $v_j \in V_{i}(\mathcal X)$, let $r_{ij}$ denote the rank of candidate $C_i$ on $v_j$'s ballot and let $m-r_{ij}$ be the Borda points associated with that rank. 
 Cham-Cour selects the committee $X$ of size $k$ which maximizes the value \[\displaystyle\sum_{C_i \in \mathcal X}\displaystyle\sum_{v_j \in V_i(\mathcal X)} m-r_{ij}.\]

\item \textbf{Expanding Approvals Rule (EAR)}:  a voting method recently created by \cite{AL}, EAR generalizes the Bucklin method to the multi-winner setting. Like STV, it proceeds in rounds in which voters can ``buy'' the seat of candidates they support. In the first round, all voters approve of only their top choice. If any candidates achieve quota ($\frac{n}{k+1}$), their supporters' votes are reduced  and the candidate is elected. If no candidates are elected, voters now support their top two choices. This continues until $k$ candidates have been elected. If two candidates make quota simultaneously, the tiebreak goes to the candidate with the higher approval score. See \cite{AL} for a complete description.  Another variant of EAR uses a slightly higher quota and a different tiebreaker, but this variant gives only modestly different results from the method we use in this paper.
    
\end{itemize}


We also describe more fully how $k$-Borda, Cham-Cour and EAR are adapted for partial ballots with an optimistic model and a pessimistic model, as outlined in \cite{BFLR}. Suppose a partial ballot of voter $i$ has length $l$, so that $l<m$ candidates are ranked on the ballot. For the $k$-Borda method, under the optimistic model (OM), if candidate $c$ appears on the ballot then they receive $m-r_{ic}$ from voter $i$, where $r_{ic}$ is their rank in voter $i$'s ballot; if $c$ does not appear on the ballot then they receive $m-l-1$ points from voter $i$. Under the pessimistic model (PM), if candidate $c$ appears on the ballot then they receive $m-r_{ic}$ from voter $i$; if $c$ does not appear on the ballot then they receive zero points from voter $i$. For Cham-Cour, the two models naturally extend when assigning points to subsets. Under OM (respectively PM), a fixed committee $X$ receives $m-l-1$ points (respectively zero points) from a voter if no candidate from $X$ appears on that voter's ballot.

The EAR method also requires a convention for voters who rank fewer than $j$ candidates when the procedure reaches the $j$th round. In the optimistic version, such a voter now supports every candidate. In the pessimistic version, voters support only candidates ranked on their ballots.  At this point, the candidates with the most support are elected until the committee is filled. These optimistic and pessimistic versions are defined and discussed in greater detail in \cite{GJM}.


\section{Technical Note on the Proofs}
\label{sec:technoteproofs}

To simplify the discussion in Appendix \ref{sec:spoilers_inevitable}  and in Appendix \ref{sec:worst_case_appendix}, we introduce some new notation. For the profile shown  in Table \ref{tab:simplified} (Top), which is symmetric in the losing candidates $L_1, L_2$ and $L_3$, we use the box notation shown in Table \ref{tab:simplified} (Bottom). More generally, each box corresponds to multiple ballots types, where the candidates inside the box are permuted in all possible orders.

\begin{table}[htbp]
\centering
\begin{minipage}{\linewidth}
\centering

\caption{The profile (Top) is symmetric with respect
to the three $L_i$ candidates; the same profile using
box notation (Bottom).}
\label{tab:simplified}

\vspace{0.15in}

\begin{tabular}{cccccc}
$5$ & $5$ & $5$ & $5$ & $5$ & $5$ \\
\hline\hline
$W$ & $W$ & $W$ & $W$ & $W$ & $W$ \\
$L_1$ & $L_1$ & $L_2$ & $L_2$ & $L_3$ & $L_3$ \\
$L_2$ & $L_3$ & $L_1$ & $L_3$ & $L_1$ & $L_2$
\end{tabular}

\vspace{0.25in}

\begin{tabular}{|c|}
\hline
$5$ \\
\hline\hline
$W$ \\
\hline
$L_i$ \\
$L_j$ \\
\hline
\end{tabular}

\end{minipage}
\end{table}

\section{Reasonable Voting Methods are Susceptible to the Spoiler Effect}\label{sec:spoilers_inevitable}

We show that most multiwinner voting methods will be susceptible to the spoiler effect by examining the election in Table \ref{tab:spoiler_inevitable}, which uses the box notation introduced in Appendix \ref{sec:technoteproofs}. 

\begin{table}[h]
    \centering
    \begin{tabular}{cccccc}
        $N$ & $N$ & $N$ & $M$ & $M$ & $M$ \\
        \hline \hline
        \cmidrule{1-3}
        \multicolumn{1}{|c|}{$W_1$} & \multicolumn{1}{|c|}{$W_1$} & \multicolumn{1}{|c|}{$W_1$} & $C_k$ & $C_{k+1}$ & $C_{k+2}$ \\
        
        \multicolumn{1}{|c|}{$\vdots$} & \multicolumn{1}{|c|}{$\vdots$} & \multicolumn{1}{|c|}{$\vdots$} & $C_{k+1}$ & $C_{k+2}$ & $C_k$ \\
        
        \multicolumn{1}{|c|}{$W_{k-1}$} & \multicolumn{1}{|c|}{$W_{k-1}$} & \multicolumn{1}{|c|}{$W_{k-1}$} & $C_{k+2}$ & $C_k$ & $C_{k+1}$ \\
        \cmidrule{1-6} 
        
        $C_k$ & $C_{k+1}$ & $C_{k+2}$ & \multicolumn{1}{|c|}{$W_1$} & \multicolumn{1}{|c|}{$W_1$} & \multicolumn{1}{|c|}{$W_1$}\\
        
        $C_{k+1}$ & $C_{k+2}$ & $C_k$ & \multicolumn{1}{|c|}{$\vdots$} & \multicolumn{1}{|c|}{$\vdots$} & \multicolumn{1}{|c|}{$\vdots$} \\
        
        $C_{k+2}$ & $C_k$ & $C_{k+1}$ & \multicolumn{1}{|c|}{$W_{k-1}$} & \multicolumn{1}{|c|}{$W_{k-1}$} & \multicolumn{1}{|c|}{$W_{k-1}$} \\
        \cmidrule{4-6}
    \end{tabular}
    \caption{A profile in which a spoiler candidate is inevitable for any reasonable multiwinner voting method when $N >> M$.}
    \label{tab:spoiler_inevitable}
\end{table}

Fix $k\ge 2$ and suppose the election contains $k+2$ candidates.  
Partition the candidates into two groups: $\mathcal{S}=\{W_1, \dots, W_{k-1}\}$ and $\mathcal{L}=\{C_k, C_{k+1}, C_{k+2}\}$. 

Let $N >> M$ and consider the preference profile in Table \ref{tab:spoiler_inevitable}. The vast majority of voters prefer   each of the candidates (equally) in $\mathcal S$  over the candidates in $\mathcal L$. If $N$ is large enough so that $\frac{3N}{k-1} > 3M$,  
any proportional method satisfying PSC must elect all candidates in $\mathcal{S}$. Similarly, any majoritarian method would also elect all of these candidates. 

The remaining voters rank every candidate in $\mathcal L$ above every candidate in $\mathcal S$. Within each of these three groups, the candidates in $\mathcal{S}$ are distributed uniformly over all possible permutations of the remaining $k-1$ positions.
In addition, the restriction of the profile to the three candidates in $\mathcal{L}$ forms a Condorcet cycle: Candidate $C_k$ defeats $C_{k+1}$, candidate $C_{k+1}$ defeats $C_{k+2}$, and candidate $C_{k+2}$ defeats $C_k$.

Any reasonable voting method elects all candidates from $\mathcal{S}$, and must select one candidate from $\mathcal{L}$. Without loss of generality, suppose the one winner from $\mathcal{L}$ is $C_k$. If we remove the losing candidate $C_{k+1}$ from the profile then $C_{k+2}$ defeats $C_k$ head-to-head by a approximately a two-to-one margin. Moreover, after $C_{k+1}$ is removed, these same margins hold for first-place support among the two remaining weak candidates. Thus, within each of the two voter blocs, $C_{k+2}$ has approximately twice as much support as $C_k$ by either measure. Again, any reasonable voting method would therefore award the final seat to $C_{k+2}$ rather than $C_k$, while continuing to elect all $k-1$ candidates in $\mathcal{S}$. Hence, the removal of the losing candidate $C_{k+1}$ would change the winning committee, and $C_{k+1}$ is a spoiler.

\section{Proportionality Axioms}\label{appendix:proportionality}

We define the axioms in terms of the Hare quota $n/k$, but each can also be defined in terms of the integer Droop quota $\lfloor n/(k+1)\rfloor +1$. The first axiom is Dummett's classic proportionality for solid coalitions, which informally says that if a voter bloc votes cohesively (i.e., the voters agree on the candidates ranked at the tops of their ballots) then the bloc should be guaranteed representation in proportion to its size.

\begin{definition}[Proportionality for Solid Coalitions (PSC), \cite{D}] 
A group of voters $N'\subseteq N$ forms a \emph{solid coalition} for a
set of candidates $C'\subseteq C$ if every voter in $N'$ ranks every
candidate in $C'$ above every candidate in $C\setminus C'$.

A committee $W$ satisfies \emph{PSC} if, for every integer
$\ell\in\{1,\ldots,k\}$, every $\ell$-large solid coalition $N'$ for a
set $C'$ receives at least
\[
    \min\{\ell,|C'|\}
\]
representatives from $C'$.  That is,
\[
    |W\cap C'| \geq \min\{\ell,|C'|\}.
\]
\end{definition}

The second axiom is local stability, which informally states that there should not be a large group of voters who all prefer the same unelected candidate to every candidate in the winning committee.

\begin{definition}[Local stability, \cite{AEFLS}]
A committee $W$ satisfies \emph{local stability} if there is no
unelected candidate $c\in C\setminus W$ and no group of voters
$N'\subseteq N$ with
\[
    |N'| \geq \frac{n}{k}
\]
such that every voter in $N'$ strictly prefers $c$ to every member of
$W$.  Equivalently, $W$ violates local stability if there exist
$c\notin W$ and a $1$-large group $N'$ such that
\[
    c \succ_i w
    \qquad
    \text{for every $i\in N'$ and every $w\in W$.}
\]
\end{definition}

The final axiom is Rank-EJR+, which informally states that if a group large enough to deserve $l$ seats all rank some candidate among their top $r$ choices, then at least one voter in that group should rank at least $l$ members of the winning committee among their top $r$ choices.

For a voter $i$ and rank threshold $r$, let
\[
    A_i(r)
    =
    \{c\in C : \operatorname{rank}_i(c)\leq r\}
\]
denote the set of candidates ranked among voter $i$'s top $r$
positions.

\begin{definition}[Rank-EJR+, \cite{BP}]
A committee $W$ satisfies \emph{rank-EJR+} if there do not exist an
integer $\ell\in\{1,\ldots,k\}$, a rank threshold $r$, an unelected
candidate $c\in C\setminus W$, and an $\ell$-large group of voters
$N'\subseteq N$ such that
\[
    c\in A_i(r)
    \qquad
    \text{for every $i\in N'$},
\]
and
\[
    |W\cap A_i(r)| < \ell
    \qquad
    \text{for every $i\in N'$.}
\]
\end{definition}

In total, we consider six notions of proportionality, as each of the three axioms above can be defined in terms of the Hare quota or the integer Droop quota, where the latter is more restrictive.

\begin{figure}
\includegraphics[width=\textwidth]{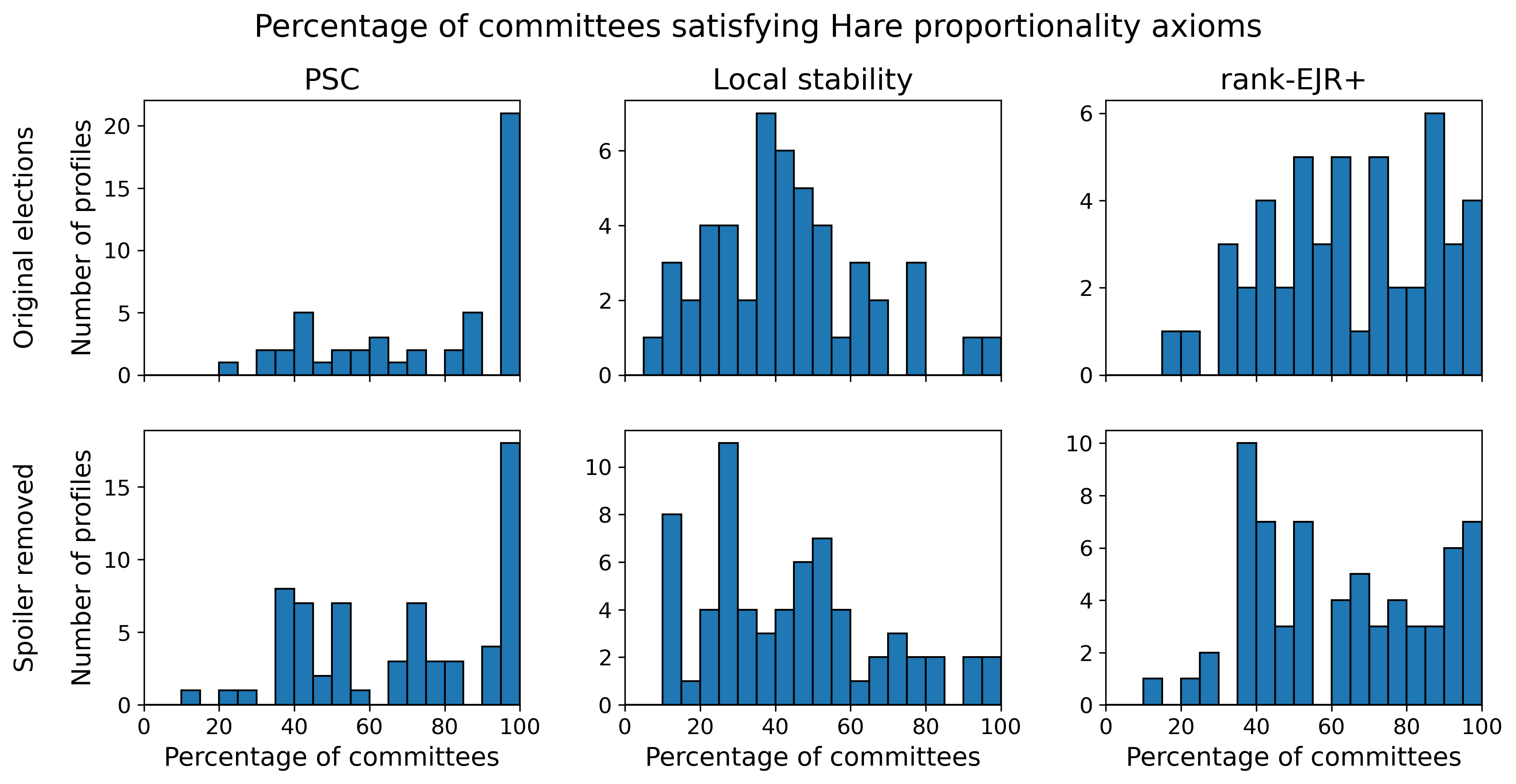}
\vspace{.1in}

\includegraphics[width=\textwidth]{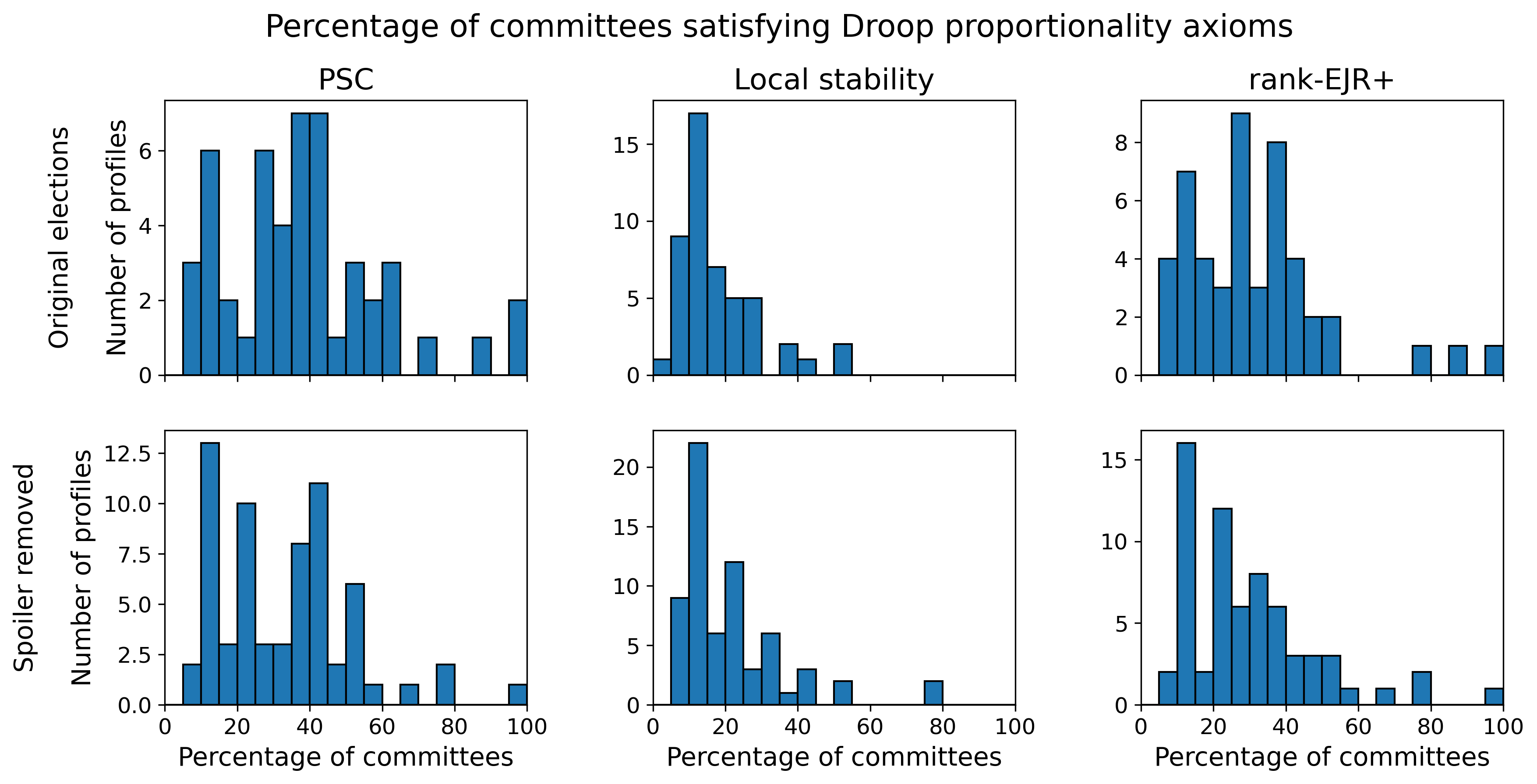}
\caption{ Histograms showing the percentage of committees of size $k$ satisfying the given axiom using the original profile or the profile with the spoiler removed. The top image uses the Hare quota and the bottom uses the integer Droop quota.}
\label{figure:proportionality_axioms}
\end{figure}

With the exception of PSC with the Hare quota, the axioms place significant restrictions on which potential winning committees are acceptable. Figure \ref{figure:proportionality_axioms} shows, for each axiom, the percentage of all possible $k$-member committees satisfying the condition. For example, across all original profiles demonstrating a spoiler effect, 21 have the property that 100\% of committees of size $k$ are compatible with Hare-PSC, showing that PSC based on the Hare quota is a weak axiom for these elections. By contrast, for local stability (based on either Hare or Droop quota), in the majority of profiles less than half of committees of size $k$ are compatible.

In many elections, a large proportion of possible committees violate the conditions. (The exception is Hare-PSC, which is well-known to be a non-restrictive axiom on the Scottish dataset \cite{BBMP24}.) Thus, the fact that both $\mathcal{W}$ and $\mathcal{W}^{-S}$ always satisfy each axiom on both the original and modified profiles is not a result of the axioms being weak or not restrictive.

\section{Worst-Case Results}\label{sec:worst_case_appendix}

In several of these proofs, we use the box notation that is introduced in Appendix \ref{sec:technoteproofs}.

\subsection{STV, SRCV, and Top-$k$ IRV}

In this section we use ``STV'' to refer to Meek or Scottish STV, as the proof applies to both methods.

\begin{lemma}\label{no_worst_case}
    Let $\mathcal{P}$ be a profile and $C$ be the first candidate eliminated by STV, SRCV, or Top-$k$ IRV. Then $C$ is not an extreme spoiler.
\end{lemma}

\begin{proof}
First, consider STV. We observe that if a candidate $C$ is removed from the voter set, the quota in $\mathcal{P}^{-C}$ is less than or equal to the quota in $\mathcal{P}$, because any bullet votes for $C$ have been removed.

Suppose, first, that some candidate $W \ne C$  is elected in $\mathcal{P}$ in the first round before $C$ is eliminated. Since $W$ must achieve quota in the profile $\mathcal{P}^{-C}$, as well as in $\mathcal{P}$,  $W$ is in the winning set for both profiles $\mathcal{P}$ and $\mathcal{P}^{-C}$. Hence $C$ cannot be an extreme spoiler. 

Suppose now that $C$ is eliminated in the first round. The STV algorithms for profiles $\mathcal P$ and $\mathcal{P}^{-C}$ then proceed in parallel, with all candidates having the same first-place totals in $\mathcal{P}$ and $\mathcal{P}^{-C}$, except that $\mathcal{P}^{-C}$, is one round ahead because $C$ did not need to be eliminated in the first round. The two profiles can only diverge once a candidate is elected, because they have different quotas. Suppose a candidate $W$ meets quota in $\mathcal{P}^{-C}$ (but not necessarily in $\mathcal{P}$).  Let $|C|$ be the number of bullet votes for $C$
and $|W|$ be the first-place votes for $W$ in the round in which it is elected in $\mathcal{P}^{-C}$ so that $|W| \geq \lfloor \frac{n-|C|}{k+1} \rfloor + 1 \geq \frac{n-|C|}{k+1}$.

We claim $W$ must also be a winner in $\mathcal P$. Suppose not, and there are $k$ other candidates that make quota or earn more votes than $W$ in the $\mathcal P$ election. There are at most $n - |C| - |W|$ first-place votes available to them. But the weakest of these candidates can receive at most 
$$\frac{n - |W| - |C|}{k} \le \frac{n- \frac{n-|C|}{k+1} - |C|}{k} = \frac{\frac{k}{k+1}(n-|C|)}{k} = \frac{n-|C|}{k+1}$$
votes, which is less than or equal to $|W|$, contradicting the claim that $W$ does not win a seat in $P$. Therefore, $w$ must win a seat in both $P$ and $P^{-C}$, so $C$ is not an extreme spoiler. 
 
Similar proofs show that $C$ cannot be an extreme spoiler under SRCV or top-$k$ IRV either.\end{proof}

Lemma \ref{no_worst_case} shows it is not possible for all losing candidates to be extreme spoilers under STV, SRCV or Top-$k$ IRV. In the next proposition, we show that there is a profile in which all losing candidates but one are extreme spoilers.

\begin{proposition} There exists a profile in which under STV, SRCV, or Top-$k$ IRV, every non-winning candidate except the first candidate eliminated is an extreme spoiler.
\end{proposition}
\begin{proof}
Consider the  profile  in Table \ref{tab:stv} with $6k^2+k$ voters and  $2k+1$ candidates denoted $W_1, \ldots, W_k$ (the winning candidates), $L$ (the first candidate eliminated), and $L_1, \ldots, L_k$ (the remaining losing candidates).  Note that the quota is $\lfloor \frac{6k^2+k}{k+1}\rfloor+1 \ge 6k-3$ for all $k \ge 2.$
\begin{table}[h]
    \centering
    \begin{tabular}{ccccc|c|ccccc|c}
        $3k$ & $k+1$ & 2 & $\dots$ & 2 & $\dots$ & $3k$ & $k+1$ & 2 & \dots & 2 & $2k$ \\
        \hline \hline
        $W_1$ & $L_1$ & $L_1$ & $\dots$ & $L_1$ & $\dots$ & $W_k$ & $L_k$ & $L_k$ & $\dots$ & $L_k$ & $L$ \\
         & $L$ & $L_2$ & $\dots$ & $L_k$ & $\dots$ & & $L$ & $L_1$ & $\dots$ & $L_{k-1}$ & 
    \end{tabular}
    \caption{Worst case STV/SRCV/Top-$k$ IRV profile.}
    \label{tab:stv}
\end{table}

Under Meek or Scottish STV, the election occurs as shown in Table \ref{tab:STV_outcomes} (Top). 
No candidate meets quota, and so the weakest candidates $L, L_1, \ldots, L_k$ are eliminated, leaving $W_1, \ldots, W_k$ as winners. 

If one of the candidates $L_j$, $j=1, \ldots k$ is removed, the election occurs as shown in Table \ref{tab:STV_outcomes} (Bottom). Note that the quota remains the same; hence no candidate meets quota and the candidates $W_1, \ldots, W_k$ are eliminated, leaving $L_1, \ldots, W_k$ as winners. 

\begin{table}[h]
    \centering
    \begin{tabular}{c|ccc}
    \hline
    Candidate & \multicolumn{3}{c}{Votes by Round}\\
        & 1 & 2 &3\\
    \hline \hline
    $W_i$ & $3k$ & $3k$ & \bm{$3k$}\\
    $L_i$ & $3k-1$ & $3k-1$ & \\
    $L$ & $2k$ & &
    \end{tabular}

    \vspace{0.25 in}

    \begin{tabular}{c|cc}
    \hline
    Candidate & \multicolumn{2}{c}{Votes by Round}\\
    \hline
    $W_i$ & $3k$ & \\
    $L_i$ & $3k+1$ & \bm{$3k+1$}  \\
    $L$ & $3k+1$ & \bm{$3k+1$}
    \end{tabular}  
    \caption{Counts by round in the original profile (Top) and in the profile after a candidate $L_j$ drops out (Bottom).}
    \label{tab:STV_outcomes}
\end{table}

Under SRCV, in the original election, the $W_i$ candidates are elected one by one with ballot totals just as the STV case. In the election where one of the candidates $L_j$  drops out, candidates $L$ and $L_i$  are elected one at a time. (As each loser candidate is elected, the other loser candidates' vote totals increase, and they each win seats.)

Finally, under  Top-$k$ IRV, in the same original  election,  the winners are  $W_1, \ldots, W_k$, since they are tied for first-place in round 1.   If $L_j$ is removed, the election proceeds as in Table \ref{tab:STV_outcomes} (Bottom). The winners in $\mathcal{P}^{-L_j}$ are all the remaining candidates since they are now tied for first-place votes in round 1.\end{proof}

\subsection{Bloc}

\begin{proposition} There exists a profile in which under Bloc, every non-winning candidate is an extreme spoiler.
\end{proposition}
\begin{proof}
Consider the profile shown in Table \ref{tab:bloc}. Candidates $W_1, \ldots, W_k$ each receive $2k+1$ votes; candidates $L_1, \ldots, L_{k+1}$ each receive $2k$ (since each of these candidates appears in  the top $k$ ranking  in $k$ columns)  and lose. However, if one of the $L_j$ candidates drops out,  then each $L_i$ gets $2(k+1) = 2k+2$ votes, and the $k$ remaining $L_i$ candidates win seats.
\begin{table}[h]
    \centering
    \begin{tabular}{c|cccc}
        $2k+1$ & 2 & 2 & $\dots$ & 2 \\
        \hline \hline
        $W_1$ & $L_2$ & $L_1$ & & $L_1$ \\
         & & $L_3$ & & \\
        $\vdots$ & $\vdots$ & $\vdots$ & $\vdots$ & $\vdots$ \\
        $W_k$ & $L_{k+1}$ & $L_{k+1}$ & & $L_k$ \\
        \hline
        $L_1$ & $L_1$ & $L_2$ & & $L_{k+1}$ \\
         & $W_1$ & $W_1$ & & $W_1$ \\
        $\vdots$ & $\vdots$ & $\vdots$ & $\vdots$ & $\vdots$ \\
        $L_{k+1}$ & $W_k$ & $W_k$ & & $W_k$ 
        
    \end{tabular}
    \caption{A profile in which every non-winning candidate is an extreme spoiler under Bloc.}
    \label{tab:bloc}
\end{table}

\end{proof}


        
    
    
        

\subsection{k-Borda}

The $k$-Borda method  is also susceptible to  the worst possible outcome, but it occurs in  different profiles for the optimistic model (OM) and pessimistic model (PM). Recall that with $k$ winners, each first-place ranking receives $k-1$ points, each second-place ranking receives $k-2$ points, and so on.

\begin{proposition} There exists a profile in which under $k$-Borda (OM), every non-winning candidate is an extreme spoiler.
\end{proposition}
\begin{proof}
Consider the profile shown in Table \ref{tab:borda_OM} where each of the candidates $W_1, \ldots, W_k$ is first-ranked in $k(k+1)$ ballots and each of the candidates $L_1, \ldots, L_{k+1}$ is first-ranked in $\frac{5}{2}k$ ballots.

\begin{table}[h]
    \centering
    \begin{tabular}{cccccc}
        1 & $\dots$ & 1 & $\frac{5}{2}k$ & $\dots$ & $\frac{5}{2}k$ \\
        \hline \hline
        $W_1$ &  & $W_k$ & $L_1$ & $\dots$ & $L_{k+1}$ \\
        \cmidrule{1-1} \cmidrule{3-3}
        \multicolumn{1}{|c|}{$L_i$} & $\dots$ & \multicolumn{1}{|c|}{$L_i$} & & & \\
        \multicolumn{1}{|c|}{$L_j$} & & \multicolumn{1}{|c|}{$L_j$} & & & \\
        \cmidrule{1-1} \cmidrule{3-3}
        
    \end{tabular}
    \caption{A profile in which every non-winning candidate is an extreme spoiler under $k$-Borda (OM).}
    \label{tab:borda_OM}
\end{table}

In the optimistic model, any candidate not appearing on a ballot receives the number of points corresponding to the next unranked spot. 
Thus, any candidate not appearing in one of the $k$ left-most types of ballots receives $2k-3$ points, and any  candidate not appearing in one of the $k+1$ right-most columns receives $2k-1$ points.
Thus, any candidate not appearing in any of the ballots receives a base number of points equal to 
$$\text{base} = [(2k-3)(k(k+1))]k + [(2k-1)\frac{5}{2}k](k+1). $$

Each $W_i$ candidate receives an extra three points 
for each of the ballots on which they are ranked first (three places higher than the first unranked spot). Thus they receive a total number of points equal to

$$W_i \text{ score} = \text{base} + 3k(k+1)=\text{base} +3k^2+3k.$$

Each candidate $L_i$ receives an additional one point for any ballot they appear on among the $k+1$ right-most columns. Additionally,  they receive an additional two points for the $k^2$ ballots on which they are ranked second and an additional one point for the $k^2$ ballots on which they are ranked third.
Thus, they receive a total number of points equal to

$$L_i \text{ score} = \text{base} + 2k^2+k^2+  \frac{5}{2}k = \text{ score} = \text{base} + 3k^2+ \frac{5}{2}k.$$
Therefore, in the original election, the $W_i$ candidates win.

Now suppose, WLOG, that  $L_1$  is removed and consider the effect on  the ballots in the left-most column of the profile in Table \ref{tab:borda_OM}.    Of the original $k(k+1)$ rankings,  $k$ rankings  had $L_1$ ranked second and $k$ rankings  had $L_1$ ranked third. Thus $2k$ rankings will now only include two candidates, and any candidate not appearing  on these ballots will receive $k-3$ points. The remaining $k(k-1)$ ballots will still have three candidates ranked; so any candidate not appearing on  these ballots will receive $k-4$ points. In addition,  among the right-most columns, there will be $\frac{5}{2}k$ ballots which are empty, so that any candidate not appearing on that ballot receives $k-1$ points. Thus, the new base value is

$$\text{base} = [(k-3)2k+(k-4)k(k-1)]k + (k-1)\frac{5}{2}k+[(k-2)\frac{5}{2}k]k.$$

In this case, the new points totals for $W_i$ and the remaining $L_i$ are 
$$W_i \text{ score} = \text{base} + 2(2k)+ 3(k(k-1)) = \text{base}+3k^2+k$$
and
$$L_i \text{ score} = \text{base} + (2(k-1)+k+1)k + \frac{5}{2}k= \text{base}+3k^2+\frac{3}{2}k.$$

Hence the remaining $L_i$ candidates win seats.

\end{proof}

\begin{proposition} There exists a profile in which under $k$-Borda (PM), every non-winning candidate is an extreme spoiler.
\end{proposition}
\begin{proof}

Consider the profile in Table \ref{tab:borda_PM}. 

\begin{table}[h]
    \centering
    \begin{tabular}{cccc}
        $x$ & $\dots$ & $x$ & $y$\\
        \hline \hline
        \cmidrule{4-4}
        $W_1$ & $\dots$ & $W_k$ & \multicolumn{1}{|c|}{$L_i$} \\
         & & & \multicolumn{1}{|c|}{$L_j$} \\
        \cmidrule{4-4}
    \end{tabular}
    \caption{A profile in which every non-winning candidate is an extreme spoiler under $k$-Borda (PM).}
    \label{tab:borda_PM}
\end{table}

In the pessimistic model, a candidate receives no points if they do not appear  on a ballot. Thus, in the original election, where each $L_i$ is first-ranked in $ky$ ballots and second-ranked in $ky$ ballots, the candidates $W_i$ and $L_i$ each receive a total number of points equal to

$$W_i \text{ score} = (2k)x$$
and 
$$L_i \text{ score} = (2k)ky+(2k-1)ky$$

On the other hand, if one of the $L_i$ is removed, each of the remaining $L_i$ candidates is first-ranked on $(k+1)y$ ballots and second-ranked on $(k-1)y$ ballots. Thus candidates $W_i$  and $L_i$ receive a total number of points equal to

$$W_i \text{ score} = (2k-1)x$$
and 
$$L_i \text{ score} = (2k-1)(k+1)y + (2k-2)(k-1)y.$$

Letting $x = 8k^3 - 5k^2 + \frac{3}{2}k$ and $y = 4k^2-2k = 2k(2k-1)$ ensures that the $W_i$ win in the original election and the $L_i$ win in the second election.

\end{proof}

\subsection{SNTV}

\vskip 2mm

\begin{proposition} There exists a profile in which under SNTV, every non-winning candidate is an extreme spoiler.
\end{proposition}
\begin{proof} Consider the profile shown in Table \ref{tab:sntv}. 
Candidates $W_i$ each receive  $2k+1$ first-place votes;  candidates $L_i$ each receive $2k$ votes and lose the election. If one of the $L_j$ candidates drops out, the remaining  $L_i$ candidates each get $2(k+1)$ votes and win the election.
\begin{table}[h]
    \centering
    \begin{tabular}{cccc}
        $2k+1$ & $\dots$ & $2k+1$ & $2$\\
        \hline \hline
        \cmidrule{4-4}
        $W_1$ & $\dots$ & $W_k$ & \multicolumn{1}{|c|}{$L_i$} \\
         & & & \multicolumn{1}{|c|}{$L_j$} \\
        \cmidrule{4-4}
    \end{tabular}
    \caption{A profile in which every non-winning candidate is an extreme spoiler under SNTV.}
    \label{tab:sntv}
\end{table}

\end{proof}

\subsection{Chamberlin-Courant}

Recall the definition of the  Chamberlin-Courant method used in this article.  For each potential winning committee $S$, assign for each voter, a number of points equal to the Borda score of the highest-ranked candidate from $S$ on that voter's ballot. For instance, if the voter's highest-ranked candidate from $S$ is second-ranked on the voter's ballot,  and if there are $2k+1$ candidates (as in the Proposition below), the committee is awarded $2k-1$ points. Under the optimistic model, if no candidate from the committee is included in the voter's ballot, the committee is awarded a number of points equal to the next unranked spot on the voter's ballot. Under the pessimistic model, if no candidate from the committee is included in the voter's ballot, the committee is awarded 0 points. In either case, the committee $S$ with the highest total score is the  Chamberlin-Courant winner set.

\begin{proposition} There exists a profile in which under Chamberlin-Courant (OM) or (PM), every non-winning candidate is an extreme spoiler.
\end{proposition}
\begin{proof}
Consider the profile shown in Table \ref{tab:CC} where $x = (k^2 + 2k)(k^2+k) = k^4+3k^3+2k^2$ and $y = 2k^4+\frac{7}{2}k^3+\frac{1}{2}k^2$.

\begin{table}[h]
    \centering
    \begin{tabular}{ccccc}
        $x$ & $y$ & $\dots$ & $x$ & $y$\\
        \hline \hline
        \multicolumn{1}{|c|}{$L_i$} & $W_1$ & & \multicolumn{1}{|c|}{$L_i$} & $W_k$ \\
        \cmidrule{2-2} \cmidrule{5-5}
        \multicolumn{1}{|c|}{$L_j$} & \multicolumn{1}{|c|}{$L_i$} & $\dots$ & \multicolumn{1}{|c|}{$L_j$} & \multicolumn{1}{|c|}{$L_i$} \\
        \cmidrule{1-1} \cmidrule{4-4}
        $W_1$ & \multicolumn{1}{|c|}{$L_j$} & & $W_k$ & \multicolumn{1}{|c|}{$L_j$}\\
        \cmidrule{2-2} \cmidrule{5-5}
    \end{tabular}
    \caption{Worst case Chamberlin-Courant profile.}
    \label{tab:CC}
\end{table}

Notice that the only committees in which every voter has a candidate ranked on their ballot are the set of winners $\{W_i\}$ and the $k+1$ sets consisting of $k$ of the $L_i$ candidates, denoted $\{L_i\}$. Therefore, these  sets are the only contenders for the winning committee. Also, these committees will have the same scores in both the optimistic and pessimistic models since every voter ranks at least one of the $W_i$ and one of the $L_i$ on their ballots.

In the original election, the scores are

$$\{W_i\} \text{ score} = (k(k+1)(2k-2)x + k(k+1)(2k)y)k$$
$$\{L_i\} \text{ score} = ((k^2(2k)+k(2k-1))x + (k^2(2k-1)+k(2k-2))y)k$$

Substituting the values of $x$ and $y$ into these expressions, we get $\{W_i\} \text{ score} > \{L_i\} \text{ score}$. 

After removing a single losing candidate, there is only one set of losing candidates $\{L_i\}$. Now, the scores are
$$\{W_i\} \text{ score} = ((2k(2k-2)+k(k-1)(2k-3))x + k(k+1)(2k-1)y)k$$
$$\{L_i\} \text{ score} = (k(k+1)(2k-1)x + k(k+1)(2k-2)y)k,$$
and a similar algebraic calculation shows that now $\{L_i\} \text{ score} > \{W_i\} \text{ score}$.

It is left as an exercise to check that any committee containing a mix of $W_i$ and $L_i$ candidates will have a smaller score, since they will be included on fewer ballots. Thus, the winning committee in the original election is $\{W_i\}$. and in election after a losing candidate is removed is $\{L_i\}.$

\end{proof}


\subsection{Expanding Approvals Rule}

Under Expanding Approvals, if the election finishes before any truncated ballots are reached, the optimistic and pessimistic versions must agree. Therefore, the following proposition requires only one profile. In addition, while Proposition \ref{prop:earextreme} uses the max approvals tie-breaking rule to choose among multiple eligible candidates in each iteration, a similar proof exists for the rank maximality tie-breaking rule preferred in~\cite{AL} using a slightly different profile with only the ballots to the left of the vertical line in Table~\ref{tab:earextreme}.

\begin{proposition} 
\label{prop:earextreme}
For any number of seats $k \geq 2$, there exists a profile in which every non-winning candidate is an extreme spoiler under Expanding Approvals (OM) or (PM) using the max approvals tie-breaker.
\end{proposition}
\begin{proof}

\begin{table}[h]
    \centering
    \begin{tabular}{ccccc|ccc}
        $3$ & $\dots$ & $3$ & $\dots$ & $3$ & $1$ & $\dots$ & $1$ \\
        \hline \hline
        $L_1$ &         & $L_1$     &           & $L_{k+1}$ & $L_1$ &       & $L_{k+1}$ \\
        $W_1$ & $\dots$ & $W_k$     & $\dots$   & $W_k$     & $L_2$ & $\dots$ & $L_1$ \\
        $L_2$ &         & $L_{k+1}$ &           & $L_k$     & $L_3$ &       & $L_2$
    \end{tabular}
    \caption{A profile in which every non-winning candidate is an extreme spoiler under EAR using the max approvals tie-breaker.}
    \label{tab:earextreme}
\end{table}

We show how to construct such a profile with $2k + 1$ candidates; $k$ winners $(W_1 \ldots W_k)$ and $k + 1$ losers $(L_1 \ldots L_{k+1})$. For each of the $k + 1$ losers, there are $3$ copies each of $k$ ballots ranking that loser first. Each of these ballots ranks a different winner second and a different other loser third. Then, there are $k+1$ additional ballots of the form $(L_i, L_{i+1 \bmod k+1}, L_{i+2 \bmod k+1})$ for $i \in \{1, \ldots, k+1\}$, such that each loser is ranked first, second, and third on exactly one of the ballots. Thus, we have $3k(k+1) + 1(k + 1) = (3k + 1)(k + 1)$ ballots total, and a candidate must exceed the Droop quota of $3k + 1$ ballots to be elected. Table~\ref{tab:earextreme} illustrates this construction.

When only the first ranks are considered, each of the losers has exactly $3k + 1$ approvals, failing to exceed the quota. After expanding to the second ranks, each loser gains $1$ extra approval for $3k + 2$ approvals total. However, each winner gains $3k + 3$ approvals, exceeding quota with a larger number of approvals. Since no two winners appear on the same ballot, electing one of them does not reduce the weight of approvals assigned to the others. Thus, each winner will be elected one-by-one until all $k$ winners have been elected. 

Now, we consider what happens when one of the losers is removed. Without loss of generality, suppose we remove $L_{k+1}$. When only the first ranks are considered, most of the losers still have exactly $3k + 1$ approvals. However, $L_1$ has $3k + 2$ approvals due to gaining an approval from the single ballot that previously ranked $L_{k+1}$ first and $L_1$ second. In addition, each of the $k$ winners now has $3$ approvals. This results in $L_1$ being elected and the weight of each ballot ranking $L_1$ first being reduced to almost $0$.

After expanding to the second ranks, we can lower bound the number of approvals for each loser by only counting the ballots which contain at most one loser in the top two ranks. Note that for a given loser $L$, this lower bound will not change when another loser $L'$ is elected and the weights are adjusted because it does not count any ballots which contain both $L$ and $L'$. Thus each remaining loser $L_i$ for $i \in \{2, \ldots, k\}$ will have $3k$ approvals from the ballots ranking them first and $3$ approvals from the ballots which had previously ranked $L_{k+1}$ first and $L_i$ third, but now rank some winner first and $L_i$ second. This means each remaining loser has at least $3k + 3$ approvals, exceeding the quota, and that lower bound does not change when other losers are elected. On the other hand, we can upper bound the number of approvals for each winner at strictly less than $3k + 3$ approvals because each winner appears on exactly $3k + 3$ ballots total, but $3$ of those ballots had their weight reduced when $L_1$ was elected. Thus, each remaining loser will be elected one-by-one until all $k$ losers except the removed loser, $L_{k+1}$, have been elected.


\end{proof}

\subsection{Minimax Condorcet Committee}

The minimax condorcet method selects as winning committee the $k$-sized committee in which every candidate is majority-preferred to every non-winning candidate, if it exists. If not, the method finds the smallest such committee of size $k^\prime>k$ and selects from this committee the $k$ candidates $W$ with the smallest  values of   $\max_{C_i \ne W } PW(W, C_i)$ where $PW(W, C_i)$ is the number of voters who prefer $W$ to $C_i$ minus the number of voters who prefer $C_I$ to $W$.

Note that if a candidate $s$ drops out, it does not affect the numbers of voters who prefer $C_i$ to $C_j,$ for any  pair of candidates, $C_i, C_j \ne s$. Thus, the only way for a non-winning candidate $L$ to replace a winner $W$ when $s$ drops out is if the smallest Condorcet Committee is larger than $k$, $L$ achieves its unique minimum margin score against $s$, and $L$'s other margin scores are larger than all the other margin scores of $W$.
That is 
\begin{equation}\label{CCMM}
PW(L, s) < \min_{C_i \ne s} PW(W, C_i) < \min_{C_i \ne s}PW(L, C_i).\end{equation}


 We use this to construct a profile with an extreme spoiler.
 
\begin{proposition}
There exists a profile in which one non-winning candidate is an extreme spoiler.
\end{proposition}

\begin{proof}
 Given a set of winners $W_1, \ldots, W_k$, a set of non-winners $L_1, \ldots, L_k,$ and an additional non-winning candidate $s$, we construct a profile with an extreme spoiler as shown in Figure \ref{MMCC}, where a weighted arrow denotes the margin of one candidate over another. In this profile, the smallest Condorcet committee is the entire candidate set. The $k$ candidates with the smallest margins of defeat are the $W_i$ candidates, so they are the winning committee. 
\begin{figure}[h]
\begin{tikzpicture}[
    >=Stealth,
    node distance=3cm and 4cm,
    every node/.style={font=\small}
]
    \node (W1) {$W_1$};
    \node (Wk) [right=of W1] {$W_k$};
    \node (S)  [below right=1.8cm and 2.5cm of Wk] {$s$};
    \node (L1) [below=4cm of W1] {$L_1$};
    \node (Lk) [below=4cm of Wk] {$L_k$};

    \path (W1) -- (Wk) node[midway] {$\dots\dots\dots\dots\dots\dots$};
    \path (L1) -- (Lk) node[midway] {$\dots\dots\dots\dots\dots\dots$};

    \draw [<->] (W1) to[bend left=30] node[above] {$0$} (Wk);
    \draw [<->] (L1) to[bend right=30] node[below] {$0$} (Lk);

    \draw [->] (L1) -- (W1) node[midway, left] {$2$};
    \draw [->] (L1) -- (Wk) node[midway, right=0.3cm, pos=0.6] {$2$};
    \draw [->] (Lk) -- (W1) node[midway, left=0.3cm, pos=0.6] {$2$};
    \draw [->] (Lk) -- (Wk) node[midway, right] {$2$};

    \draw [->] (W1) -- (S) node[midway, below right] {$6$};
    \draw [->] (Wk) -- (S) node[midway, above right] {$6$};

    \draw [->] (S) -- (L1) node[midway, above] {$4$};
    \draw [->] (S) -- (Lk) node[midway, below right] {$4$};

\end{tikzpicture}
\caption{Margin graph of a profile with an extreme spoiler under the Minimax Condorcet method.}
\label{MMCC}
\end{figure}

If $s$ is removed, however, $\{L_1, \dots, L_k\}$ is a Condorcet committee of size $k$, so the $L_i$ candidates are elected. Therefore, $s$ is an extreme spoiler. 

Because any margin graph satisfying margin parity (in this case, all margins are even) can be realized~\cite{DHP}, there exists a preference profile with this margin graph that has an extreme spoiler.
\end{proof}

\begin{proposition}
Two extreme spoilers cannot both get the other elected by dropping out.
\end{proposition}

\begin{proof}
Let $s_1$ and $s_2$ be extreme spoilers.

Suppose that $s_2$ becomes a winner if $s_1$ is removed. Because $s_1$ is an extreme spoiler, by Equation (\ref{CCMM}), for all winners $W$, 
$$PW(s_2, s_1) < \min_{C_i \ne s} PW(W, C_i) < \min_{C_i \ne s}PW(s_2, C_i).$$
In particular, $PW(W, s_2) < PW(s_2, W)$ which implies $PW(W, s_2)<0$. It follows that $PW(s_2, s_1)<0$. 

But by a similar argument, if $s_1$ becomes a winner when $s_2$ is removed, then $PW(s_1, s_2)<0$. These two inequalities cannot both be true, so at most one of the extreme spoilers can cause the other to be elected.
\end{proof}

As a corollary to this, because two extreme spoilers cannot drop out and cause each other to be elected, there cannot be an election with only $k+1$ losers where all losers are extreme spoilers.

\section{Counting Clones that Impact Elections}\label{sec:clone_appendix}

Table \ref{clone_results_clone_count} in this appendix is similar to Table~\ref{clone_results}, but counts the number of clones, not the number of elections.

\begin{table}
\centering
{\small 
\begin{tabular}{c||ccc||ccc}
&\multicolumn{2}{c}{$L \rightarrow W, W$ }& & \multicolumn{2}{c}{$L \rightarrow W, L$} \\
\hline \hline
Method & 0\% & 10\% & 50\%  & 0\% & 10\% & 50\% \\
\hline \hline
Bloc  & 1 & 1 & 5 & 334 & 310 &136\\
$k$-Borda OM & 0 & 0 & 44 & 1222 &1187 &479\\
$k$-Borda PM & 0 & 0 & 0 & 241 & 192 & 4\\
SRCV & 0 & 0 & 0 & 0 & 0 & 0\\
MM Cond Committee & 0 & 0 & 0 &2 & 2 & 2\\
SNTV& 0 & 0 & 0 & 0 & 0 & 0\\
Top $k$ IRV & 0 & 0 & 0 &3 & 3 & 3\\
Cham-Cour OM & 0 & 0 & 0 & 1483 & 1422 &647\\
Cham-Cour PM & 0 & 0 & 0 & 848	 & 780 & 285\\
Scottish STV& 0 & 0 & 0 & 0 & 0 & 0\\
Meek STV & 0 & 0 & 0 &0 & 0 & 0\\
EAR OM & 0 & 0 & 0 & 1290 & 1245 & 669\\
EAR PM & 0 & 0 & 0 & 736 & 706 & 408\\
\hline
\end{tabular}

\vspace{0.1 in}

\begin{tabular}{c||ccc||ccc}
\hline
&\multicolumn{2}{c}{$W \rightarrow L, L$}& & \multicolumn{2}{c}{$L \rightarrow L, L+other$} \\
\hline
Method & 0\% & 10\% & 50\%  & 0\% & 10\% & 50\% \\
\hline
Bloc  & 0 & 2 & 90 & 166 & 168 & 177\\
$k$-Borda OM & 0 & 0 & 0 & 109 & 111 & 142\\
$k$-Borda PM & 0 & 0 & 109 & 44 & 46 & 51\\
SRCV & 0 & 0 & 0 & 0 & 0 & 0\\
MM Cond Committee & 0 & 0 & 0 &0 & 0 & 0\\
SNTV& 0 & 154 & 1626 & 0 & 0 & 0\\
Top $k$ IRV & 0 & 0 & 0 &0 & 0 & 0\\
Cham-Cour OM & 0 & 0 & 0 & 67 & 68 & 81\\
Cham-Cour PM & 0 & 0 & 0 &90 & 92 & 101 \\
Scottish STV& 0 & 0 & 0 & 0 & 0 & 0\\
Meek STV & 0 & 0 & 0 &0 & 0 & 0\\
EAR OM & 0 & 0 & 10 &138 & 143 & 173\\
EAR PM & 0 & 0 & 2 &82 & 83 & 90\\
\hline
\end{tabular}

\vspace{0.1 in}

\begin{tabular}{c||ccc||ccc}
\hline
&\multicolumn{2}{c}{$W \rightarrow W, L + other$}& & \multicolumn{2}{c}{$W \rightarrow W, W$} \\
\hline
Method & 0\% & 10\% & 50\%  & 0\% & 10\% & 50\% \\
\hline
Bloc  &81 & 78 & 29 &   2021 & 2054 &2361\\
$k$-Borda OM & 76	 & 61	 & 3     & 1730 & 1883 & 2790\\
$k$-Borda PM & 64 & 54 & 23    &1528 & 1606 & 2040 \\
SRCV & 0 & 0 & 0 & 2474	& 2474 & 2474\\
MM Cond Committee & 0 & 0 & 0 &2414 & 2414 & 2441\\
SNTV& 0 & 0 & 0 & 0 & 0 & 382\\
Top $k$ IRV & 0 & 0 & 0 &0 & 0 & 351\\
Cham-Cour OM & 0 & 40 & 224     & 0 & 0 & 167\\
Cham-Cour PM & 94 & 79 & 161    &0 & 0 & 8 \\
Scottish STV& 12 & 18& 45 &  207 & 207 & 289 \\
Meek STV & 3 &3 & 12   &284 & 284 &	372 \\
EAR OM & 226 & 242 & 196 &465 & 465 & 799\\
EAR PM & 92 & 103 & 119 &502 & 514 & 642\\
\hline
\end{tabular}
}
\caption{Number of clones that can be added to affect the outcome of an election.}\label{clone_results_clone_count}

\end{table}

\FloatBarrier

\section{Similarity Graphs}\label{appendix_section:graphs}

\begin{figure*}[h]
    \centering
    \includegraphics[width=0.7\linewidth]{"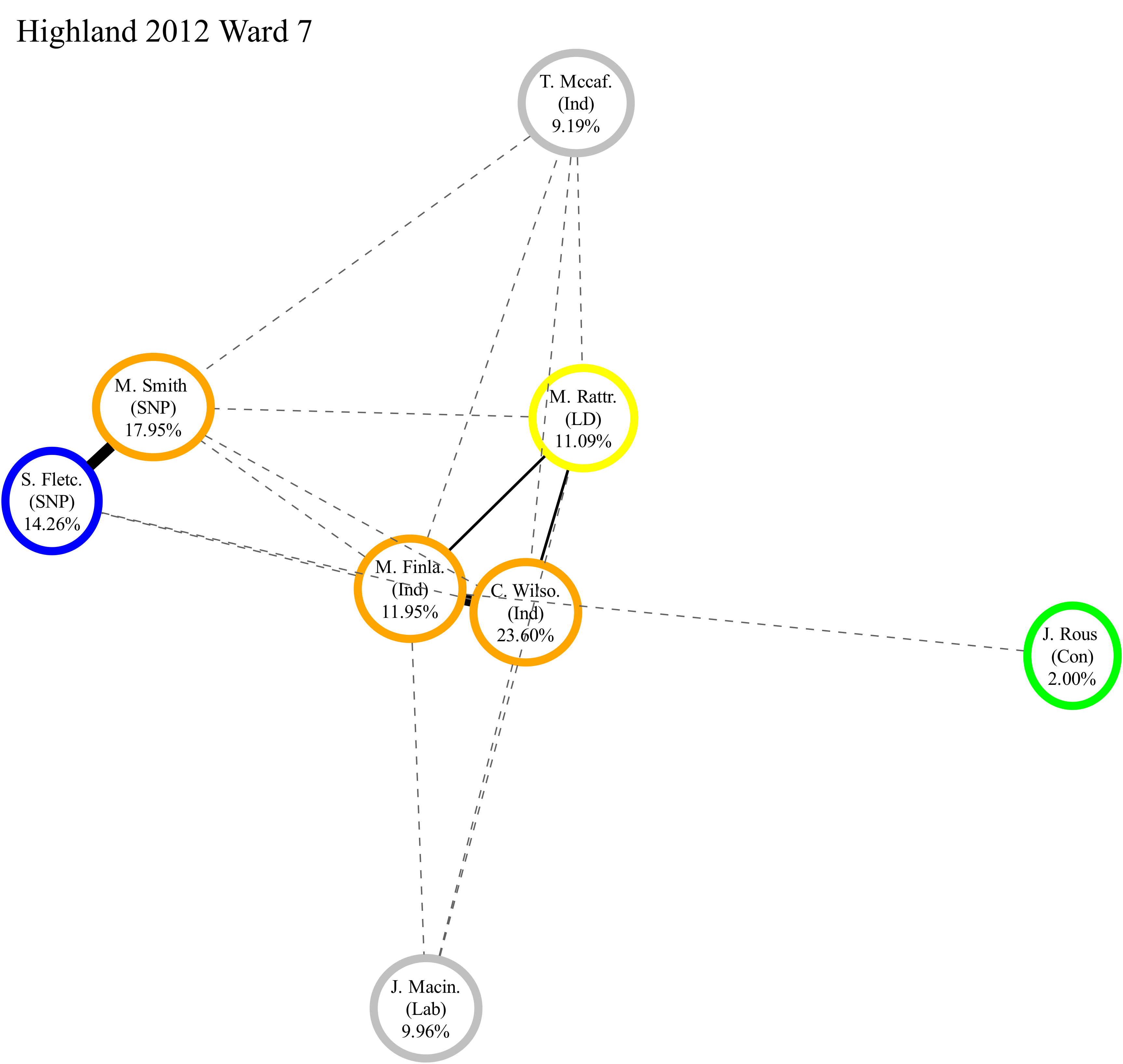"}
    \quad
    \includegraphics[width=0.25\linewidth]{"SpoilerPaperLegend.png"}
    \caption{Mutual preference graph visualizations of 2012 Highland Ward 7 election from Table \ref{table:highland_2012_ward7}. The yellow (real winner) and orange candidates won in the real-world election. However, removing the green candidates (spoilers) causes the blue candidate (alternate winner) to replace the yellow candidate in the winning set. This is the one Scottish spoiler event in which a ``fringe'' candidate is a spoiler candidate.}
    \label{fig:spoilergraphsappendix1}
\end{figure*}

\begin{figure*}[tp]
    \centering
    \includegraphics[width=0.65\linewidth]{"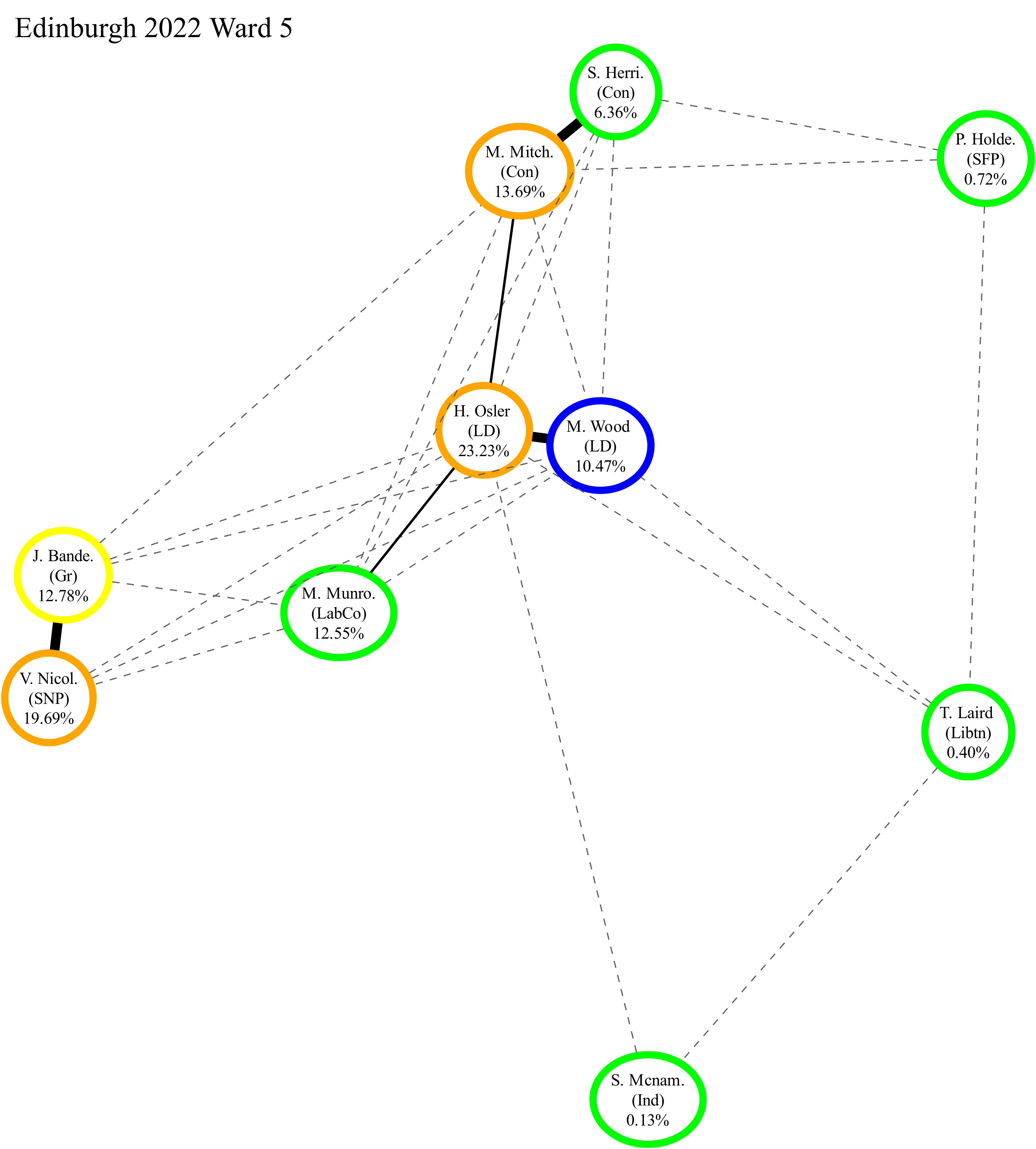"}

    \medskip
    \includegraphics[width=0.65\linewidth]{"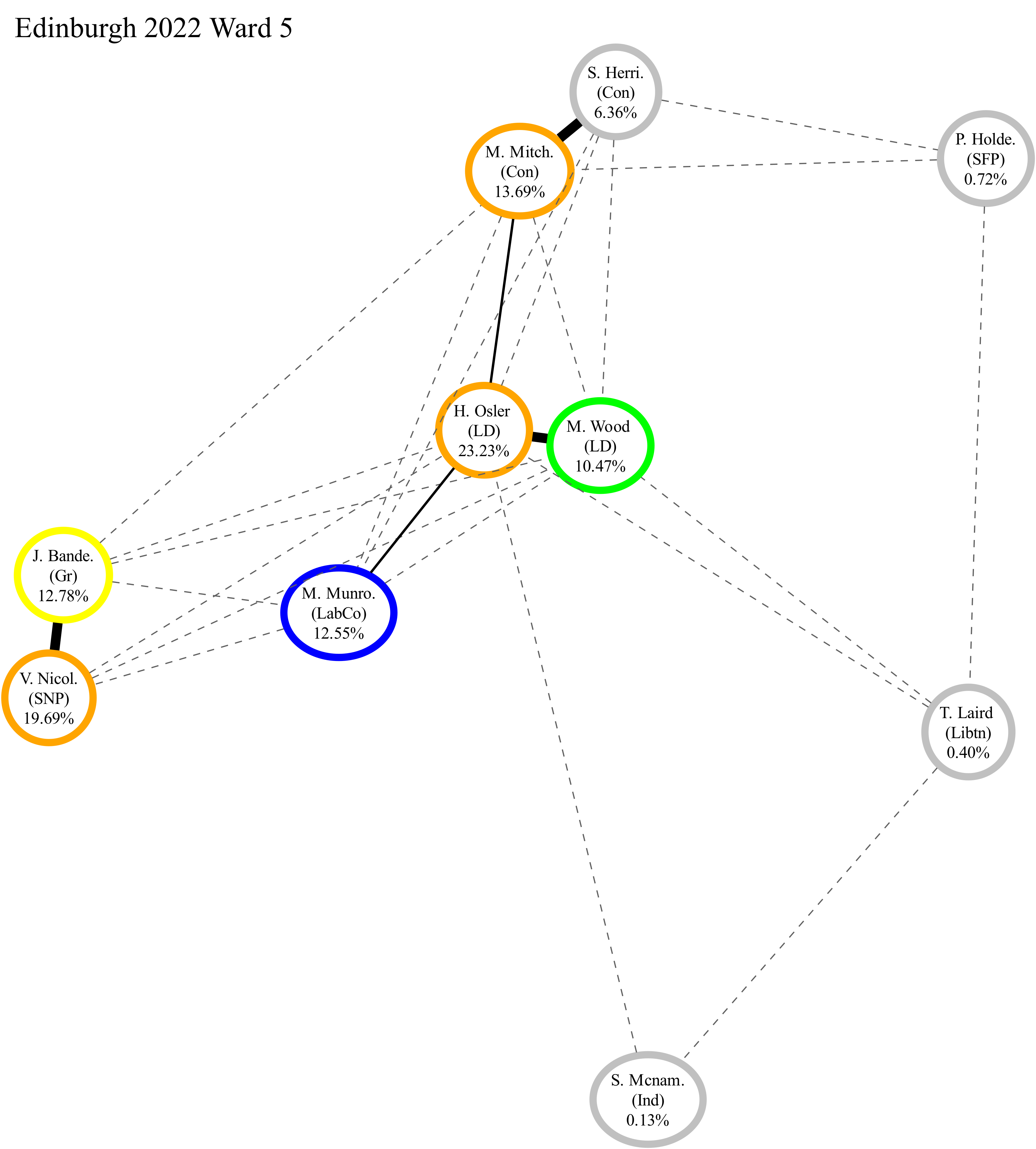"}
    \quad
    \includegraphics[width=0.25\linewidth]{"SpoilerPaperLegend.png"}
    \caption{Mutual preference graph visualizations of the Edinburgh 2022 Ward 5 election discussed in Example \ref{edinburgh}. The yellow (real winner) and orange candidates won in the real world election. However, removing one of the green candidates (spoilers) causes the blue candidate (alternate winner) to replace the yellow candidate in the winning set.}
    \label{fig:spoilergraphsappendix}
\end{figure*}

In this section, we include two figures illustrating STV elections discussed in the paper. Figure \ref{fig:spoilergraphsappendix1} illustrates the 2012 election from Highland Ward 7 shown in Table \ref{table:highland_2012_ward7}. Figure \ref{fig:spoilergraphsappendix} illustrates the 2022 election from Edinburgh Ward 5 discussed in Example \ref{edinburgh}. The top graph of \ref{fig:spoilergraphsappendix} highlights five of the spoiler candidates in this election. Removing any one of these candidates causes M. Wood to replace J. Bandel in the winner set. The bottom graph shows the sixth spoiler, M. Wood, whose removal causes Munro-Brian to replace Bandel in the winner set.

\end{document}